\documentclass[aps,prb,10pt,amsmath,amssymb,twocolumn,longbibliography,superscriptaddress]{revtex4-2}
\usepackage{hyperref}
\usepackage{graphicx}
\usepackage{mathtools}
\usepackage{physics}

\def\be#1\ee{\begin{align}#1\end{align}}

\begin{document}

\title{Polarization Jumps across Topological Phase Transitions in Two-dimensional Systems}

\author{Hiroki Yoshida}
\affiliation{Department of Physics, Tokyo Institute of Technology, 2-12-1 Ookayama, Meguro-ku, Tokyo 152-8551, Japan}
\author{Tiantian Zhang}
\affiliation{Department of Physics, Tokyo Institute of Technology, 2-12-1 Ookayama, Meguro-ku, Tokyo 152-8551, Japan}
\affiliation{Institute of Theoretical Physics, Chinese Academy of Sciences, Beijing 100190, China}
\author{Shuichi Murakami}
\affiliation{Department of Physics, Tokyo Institute of Technology, 2-12-1 Ookayama, Meguro-ku, Tokyo 152-8551, Japan}

\date{\today}

\begin{abstract}
    In topological phase transitions involving a change in topological invariants such as the Chern number and the $\mathbb{Z}_2$ topological invariant, the gap closes, and the electric polarization becomes undefined at the transition. In this paper, we show that the jump of polarization across such topological phase transitions in two dimensions is described in terms of positions and monopole charges of Weyl points in the intermediate Weyl semimetal phase. We find that the jump of polarization is described by the Weyl dipole at $\mathbb{Z}_2$ topological phase transitions and at phase transitions without any change in the value of the Chern number. Meanwhile, when the Chern number changes at the phase transition, the jump is expressed in terms of the relative positions of Weyl points measured from a reference point in the reciprocal space.
\end{abstract}

\maketitle

\section{Introduction}
\label{sec:introduction}

The notion of topology is now widely used to classify electronic states in condensed matter physics. This approach was pioneered by Thouless, Kohmoto, Nightingale, and den Nijs through the study of the integer quantum Hall effect~\cite{Klitzing1980,Thouless1982}. They showed that the quantized Hall conductance can be expressed using the TKNN integer, which is now called the Chern number. This finding has led to the theoretical discovery of topological insulators. Two-dimensional topological insulators have novel edge states~\cite{Jackiw1976} which persist unless the energy gap of the system is closed. At the gap closing, there can appear another topological phase called the topological semimetal phase such as Dirac semimetal~\cite{Wang2012,Young2012,Young2015} and Weyl semimetal~\cite{Wan2011,Huang2015,Lv2015,Xu2015} phases. In particular, Weyl semimetals have attracted much interest because of their unique properties such as the stability against symmetry-preserving perturbations in three dimensions.

Here we focus on the electric polarization in the context of topology in the electronic band structure. The electronic contribution to the polarization in crystals is described by the Berry curvature of Bloch states according to the modern theory of polarization~\cite{Resta1992,King-Smith1993,Vanderbilt1993}. The electric polarization is determined only in terms of modulo the polarization quantum and is well-defined in insulating systems. Electric polarization in topologically nontrivial phases is a focus of active study recently~\cite{Hetenyi2022,Zhang2022,Vaidya2023}. We recently discovered that the electric polarization in an insulating two-dimensional system can have a jump when the system changes across the Weyl semimetal phase~\cite{Yoshida2023}. The jump of polarization $\Delta \vb{P}$ is described by using a newly introduced quantity “Weyl dipole” representing how the Weyl points in the intermediate Weyl semimetal phase are displaced in the reciprocal space. This result is applicable to all types of Weyl semimetal phases between two normal insulator phases.

In this paper, we study the jump of polarization at topological phase transitions between two topologically distinct insulator phases. Unlike the case of the Weyl semimetal phase between two normal insulator phases in Ref.~\cite{Yoshida2023}, the two insulating phases are topologically distinct and cannot be adiabatically connected. Hence, it is not trivial how to define a jump of polarization at the transition. Furthermore, when the Chern number of the system changes, Weyl points do not necessarily appear in pairs at the topological phase transition, which hinders the construction of the Weyl dipole. In this paper, we find that by setting the reference points in the phase diagram where the polarization is constrained by symmetry, we can define the jump of polarization at topological phase transitions as well, and it can be described by positions of the Weyl points. In short, for general two-dimensional topological phase transitions, the jump of electric polarization can be described by the sum of the products of monopole charges and relative positions of Weyl points from the reference point. We note that if the Chern number remains unchanged at the transition, such as a $\mathbb{Z}_2$ topological phase transition, the jump of polarization is still given by the Weyl dipole.

This paper is organized as follows. In Sec.~\ref{sec:RevPol}, we review the modern theory of polarization and our previous findings for completeness. Then, we use the Haldane model~\cite{Haldane1988} as an example to investigate the jump at the phase transition between a Chern insulator and a normal insulator phase in Sec.~\ref{sec:Haldane}. In Sec.~\ref{sec:Kane-Mele}, we use the Kane-Mele model~\cite{Kane2005a} as an example of a $\mathbb{Z}_2$ topological insulator and discuss its electric polarization and jumps of electric polarization. In Sec.~\ref{sec:origin}, we consider the origin of jumps of polarization at topological phase transitions. In Sec.~\ref{sec:material}, we propose BaMnSb$_2$ as a candidate material to observe the jump at topological phase transitions. We conclude this paper in Sec.~\ref{sec:conclusion}.

\section{Review of modern theory of polarization and jumps of electric polarization}
\label{sec:RevPol}

We briefly review the modern theory of polarization and the basic understanding of jumps of polarization gained in Ref.~\cite{Yoshida2023} before dealing with topological phase transitions.

Consider a two-dimensional insulator with primitive lattice vectors $\vb{a}_{1}$ and $\vb{a}_{2}$. Let $u_{\vb{r},n}(\vb{k})$ denote the cell periodic part of the Bloch state of the $n$-th energy level at position $\vb{r}$. Then, the electronic polarization of this system under the wavevector-periodic gauge condition for energy eigenfunctions, $u_{\vb{r},n}(\vb{k})=e^{i\vb{b}\cdot\vb{r}}u_{\vb{r},n}(\vb{k}+\vb{b})$, is given by~\cite{King-Smith1993,Vanderbilt1993}
\be
    \vb{P} =\frac{-ie}{(2\pi)^2}\int_{\mathrm{BZ}}\mathrm{d}^2k\sum_n^{occ}\bra{u_n(\vb{k})}\pdv{}{\vb{k}}\ket{u_n(\vb{k})}\,\qty(\mathrm{mod}\frac{e}{\Omega}\vb{a}_{1,2})\label{eq:Polarization_KV},
\ee
where $\vb{b}$ is the reciprocal lattice vector, $-e$ is a charge of an electron ($e>0$), $\Omega$ is the area of the unit cell, the integral is over the Brillouin zone, and the sum is taken over all the occupied bands. In addition to the electronic contribution in Eq.~\eqref{eq:Polarization_KV}, we also have an ionic contribution. Each contribution depends on the convention of the lattice, i.e., the origin of the coordinate, but their sum does not.

\begin{figure}[t]
    \begin{center}
        \includegraphics[width=\columnwidth,pagebox=artbox]{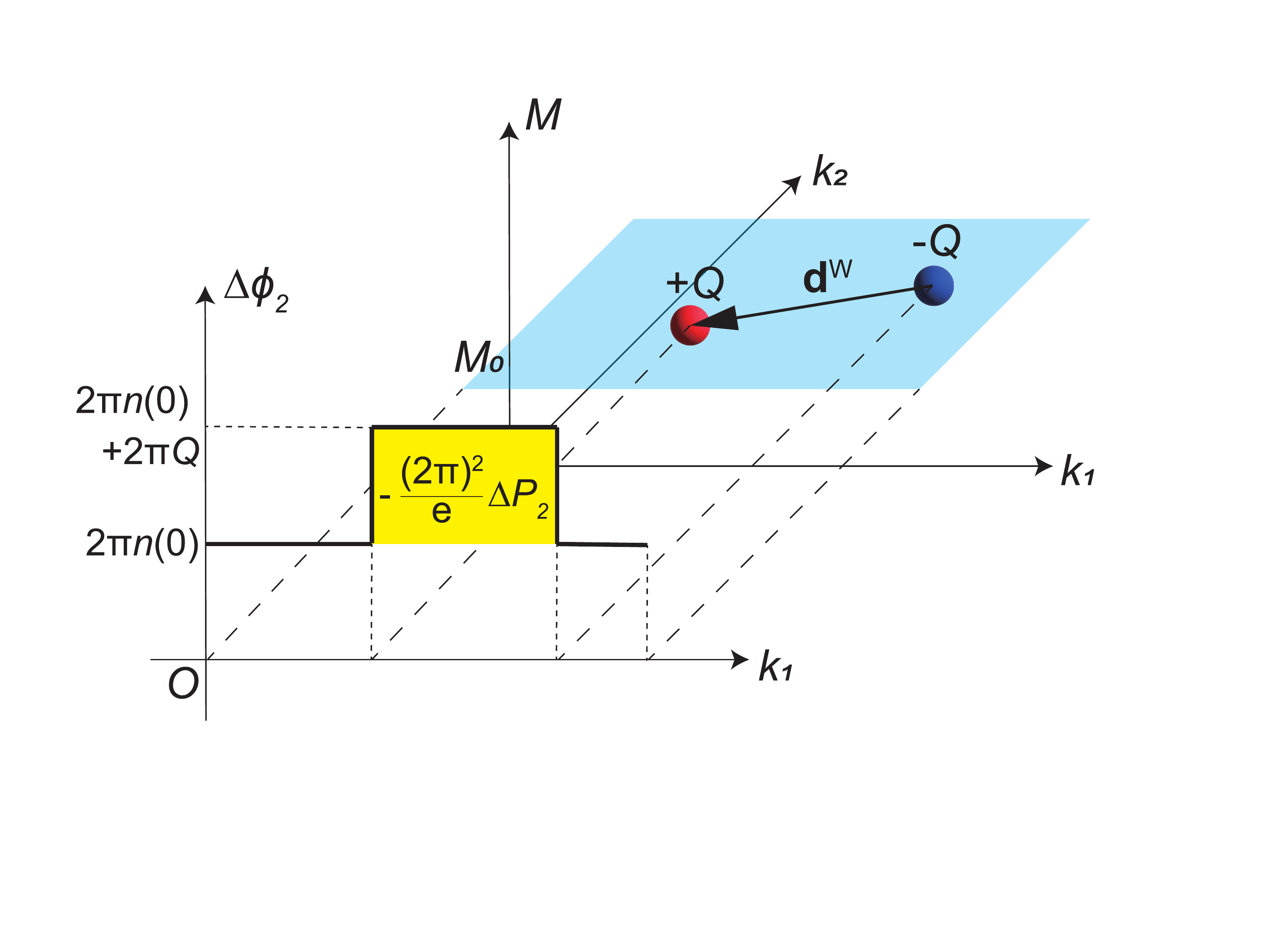}
        \caption{A pair of Weyl points with monopole charges $\pm Q$ in the $(\vb{k},M)$ space, appearing at the phase transition between two insulating phases. The difference of Berry phases $\Delta\phi_2$ in the limits $M\to M_0\pm0$ jumps by $\pm2\pi Q$ at these Weyl points. The area shown by a yellow rectangle contributes to the jump of polarization across $M=M_0$.}
        \label{fig:3DkM_normal}
    \end{center}
\end{figure}

We next review the theory of the jump of polarization~\cite{Yoshida2023}. We consider a two-dimensional system with a real parameter $M$, and we assume that the system is a Weyl semimetal when the parameter $M$ is equal to $M_0$ and otherwise becomes a normal insulator. Then, the polarization is ill-defined at $M=M_0$. Between the both sides of this value, $M = M_0\pm\delta$ ($\delta$: positive infinitesimal), the electronic polarization may have a jump, as proposed in Ref.~\cite{Yoshida2023}. In Ref.~\cite{Yoshida2023}, we found that this jump is beautifully and compactly expressed in terms of a ``Weyl dipole''. The jump of polarization along the direction of $\vb{a}_2$ between the two limits $M\to M_0\pm0$ can be written as
\be
    \Delta P_{2}&=\frac{-e}{(2\pi)^2}\int_0^{b_{1}}\mathrm{d}k_{1}\Delta\phi_{2}(k_{1})\sin\theta,\label{eq:DeltaPolarization}\\
    \Delta \phi_{2}(k_{1})&\coloneqq\phi_{2}^+(k_{1})-\phi_{2}^-(k_{1}),\label{eq:DeltaPhi}\\
    \phi_{2}^{\pm}(k_{1})&\coloneqq\lim_{M\to M_0\pm0} i\sum_{n}^{occ}\int_0^{b_{2}}\mathrm{d}k_{2}\bra{u_n(\vb{k})}\pdv{}{k_{2}}\ket{u_n(\vb{k})},\label{eq:pmPhi}
\ee
where $\vb{b}_{1}$ and $\vb{b}_{2}$ are reciprocal lattice vectors corresponding to $\vb{a}_{1}$ and $\vb{a}_2$, $b_{i} \coloneqq\abs{\vb{b}_{i}}$, $\theta$ is an angle between $\vb{b}_1$ and $\vb{b}_2$, the integration is over the parallelogram spanned by $\vb{b}_1$ and $\vb{b}_2$ instead of the Brillouin zone for convenience, and $k_1$ and $k_2$ are the components of $\vb{k}$ along $\vb{b}_1$ and $\vb{b}_2$, respectively. Here, $\Delta \phi_{2}(k_{1})$ represents a difference of Berry phases between the two limits $M\to M_0\pm0$. Since the Berry phases become the same between the two limits, the difference of Berry phases must be zero in terms of modulo $2\pi$, i.e.
\be
    \Delta \phi_{2}(k_{1})=2\pi n(k_{1})\quad(n(k_{1})\in\mathbb{Z}).\label{eq:DeltaPhi_2pin}
\ee
We also find that as a function of $k_1$, $\Delta \phi_2$ jumps by $2\pi Q$ across the projection of the Weyl point, where $Q$ is a topological quantity called monopole charge defined for Weyl points in the three-dimensional $(\vb{k},M)$ space. In Fig.~\ref{fig:3DkM_normal}, we show an example where Weyl points with monopole charges $\pm Q$ are present in the $(\vb{k},M)$ space with $\vb{b}_1\perp\vb{b}_2$ for simplicity. The corresponding $\Delta\phi_2$ as a function of $k_1$ is also plotted. Then, $\Delta P_2$ can be calculated as an area of the yellow rectangle. Similar analysis can be done for $\Delta P_1$ and the jump of polarization is expressed as
\be
    \Delta\vb{P}=\frac{e}{2\pi}\hat{z}\times\vb{p}^W\quad\qty(\mathrm{mod}\frac{e}{\Omega}\vb{a}_{1,2}),\label{eq:DeltaP_NI}
\ee
where $\hat{z}$ is a unit vector perpendicular to the two-dimensional system. Here we introduced the notion of ``Weyl dipole" $\vb{p}^W$ as
\be
    \vb{p}^W \coloneqq Q\vb{d}^W\quad(\mathrm{mod}\, Q\vb{b}_{1,2}),
\ee
where $\vb{d}^W$ is a displacement vector from a Weyl point with a monopole charge $-Q$ to the one with $+Q$. The jump is experimentally measurable through the difference of surface charge densities or the electric current that flows during the adiabatic change connecting two states before and after the jump. This result applies to jumps of polarization between two-dimensional topologically trivial insulating phases via a Weyl semimetal phase in general cases~\cite{Yoshida2023}. In such cases, Weyl points with opposite monopole charges appear in pairs at the transition at $M=M_0$ without a change of the Chern number of the system. In the following, we examine how these results are changed in the case of topological phase transitions. 

\section{Polarization jump between a Chern insulator and a normal insulator phase}
\label{sec:Haldane}

We first construct a general theory of the polarization jump between a Chern insulator and a normal insulator phase. First, we study the renowned Haldane model as an example~\cite{Haldane1988}. The original Haldane model is a tight-binding model of spinless electrons on a hexagonal lattice with staggered on-site potentials, real nearest neighbor (NN), and complex next nearest neighbor (NNN) hoppings. It has a three-fold rotational symmetry. To get a nontrivial value of the electric polarization, we need to break this symmetry. For this purpose, we here use an extended Haldane model proposed in Ref.~\cite{Coh2009}.

\subsection{Extended Haldane model}
\label{subsec:ExtendedHaldane}

\begin{figure}[t]
    \begin{center}
        \includegraphics[width=\columnwidth,pagebox=artbox]{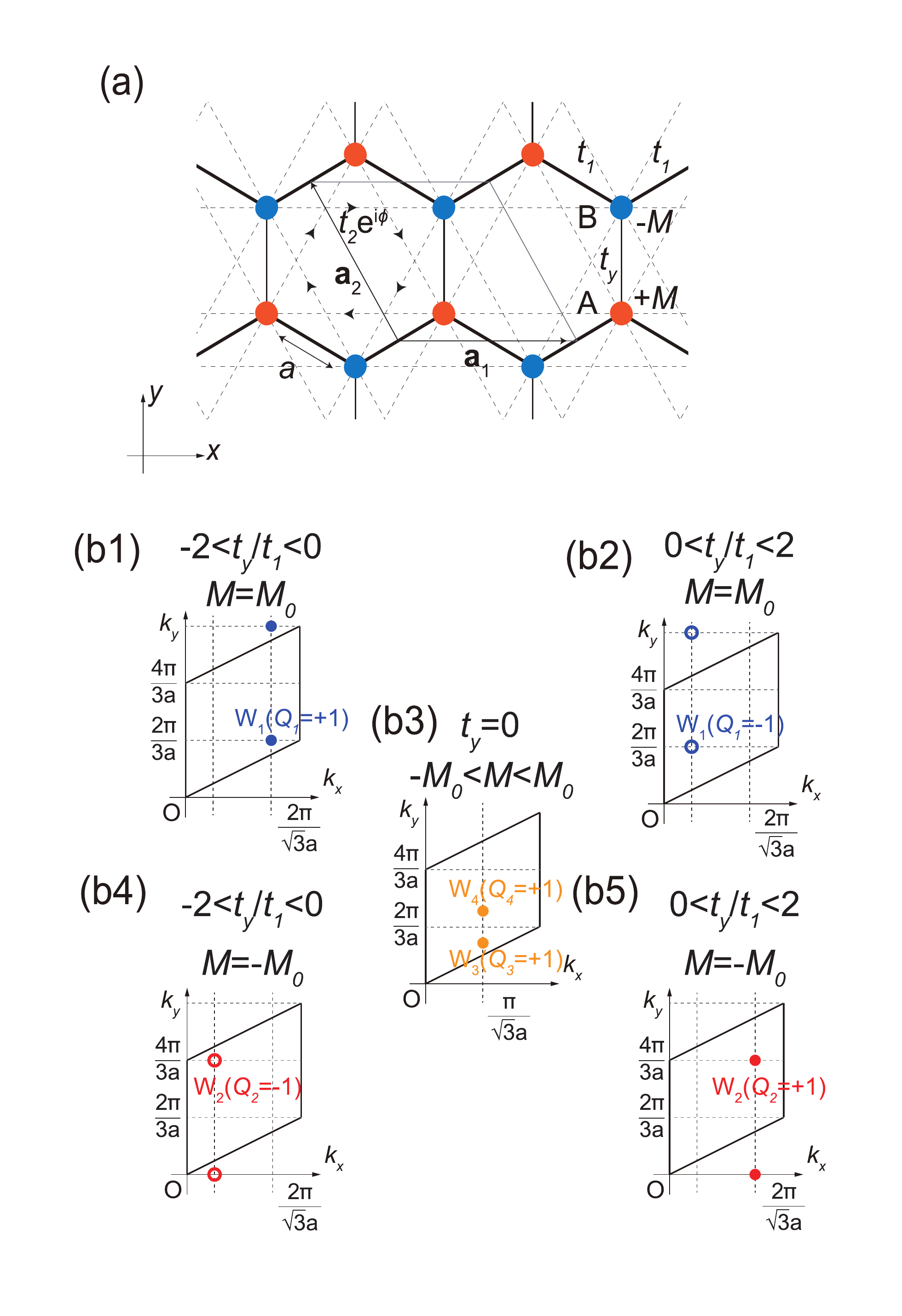}
        \caption{The extended Haldane model without $C_3$ rotational symmetry. (a) A schematic illustration of the extended Haldane model. Compared to the ordinary Haldane model, a nearest neighbor anisotropic hopping term for the NN hopping along the $y$ axis is modulated from $t_1$ to $t_y$. (b1)-(b5) The distributions of Weyl points $W_{1},\ldots,W_4$ in the reciprocal space for various values of $t_y/t_1$ and $M$. The monopole charge of each Weyl point within the three-dimensional $(\vb{k},M)$ space is also shown. We define $M_0$ as $M_0=4t_2\qty(1+t_y/(2t_1))\sqrt{1-(t_y/(2t_1))^2}$. We set $\sin\phi>0$ as an example. The black parallelograms are the unit cell of the reciprocal lattice.}
        \label{fig:ExtHaldane}
    \end{center}
\end{figure}

As in the original paper~\cite{Haldane1988}, we parametrize NN hoppings by $t_1(>0)$, NNN hoppings in the direction of arrows in Fig.~\ref{fig:ExtHaldane}(a) by $t_2e^{i\phi}\ (t_2>0,\phi\in\mathbb{R})$, and staggered on-site potentials by $+M$ for A sites and $-M$ for B sites. In the extended Haldane model proposed in Ref.~\cite{Coh2009}, we break the three-fold rotational symmetry by replacing the NN hopping parameter $t_1$ by $t_y$ for NN hoppings in the $y$-direction while keeping the NN hopping parameters to other directions to be $t_1$ (Fig.~\ref{fig:ExtHaldane}(a)). We take primitive lattice vectors to be $\vb{a}_1=a(\sqrt{3},0)$ and $\vb{a}_2=\frac{a}{2}(-\sqrt{3},3)$, where $a$ is a lattice constant. The reciprocal lattice vectors corresponding to $\vb{a}_1$ and $\vb{a}_2$ are $\vb{b}_1=\frac{2\pi}{\sqrt{3}a}\qty(1,\frac{1}{\sqrt{3}})$ and $\vb{b}_2=\frac{4\pi}{3a}\qty(0,1)$, respectively. The Hamiltonian of this system is given as
\be
    H(\vb{k}) &= g_1(\vb{k})I + \Re f(\vb{k})\sigma^x\nonumber\\
    &\qquad +\Im f(\vb{k})\sigma^y+(M+g_2(\vb{k}))\sigma^z,
\ee
where $\sigma$'s are the Pauli matrices and 
\be
    f(\vb{k})&=2t_1\cos\qty(\frac{\sqrt{3}a}{2}k_x)e^{i\frac{a}{2}k_y}+t_ye^{-iak_y},\label{eq:f}\\
    g_1(\vb{k})&=2t_2\cos\phi\sum_{i=1}^3\cos(\vb{k}\cdot\vb{d}_i),\\
    g_2(\vb{k})&=-2t_2\sin\phi\sum_{i=1}^3\sin(\vb{k}\cdot\vb{d}_i)\label{eq:g2},
\ee
for $\vb{d}_1=a(\sqrt{3},0),\,\vb{d}_2=\frac{a}{2}\qty(-\sqrt{3},3),\,\vb{d}_3=\frac{a}{2}\qty(-\sqrt{3},-3)$. Then, the energy eigenvalues of this system are given by $E_{\pm}=g_1\pm\sqrt{\abs{f}^2+(M+g_2)^2}$.

For $-2<t_y/t_1<2,\,t_y\neq 0$, this system becomes a Weyl semimetal at $M=\pm M_0\sin\phi$, where $M_0\coloneqq4t_2\qty(1+t_y/(2t_1))\sqrt{1-(t_y/(2t_1))^2}$. When $M= M_0\sin\phi$, the Weyl point is located at
\be
    W_1:\qty(k_x,k_y)&=\qty(\frac{2}{\sqrt{3}a}\cos^{-1}\qty(\frac{t_y}{2t_1}),\frac{2\pi}{3a}),\label{eq:W1}
\ee
and when $M=- M_0\sin\phi$, the Weyl point is located at
\be
    W_2:\qty(k_x,k_y)&=\qty(\frac{2}{\sqrt{3}a}\cos^{-1}\qty(-\frac{t_y}{2t_1}),\frac{4\pi}{3a})\label{eq:W2},
\ee
where we set $0\leq \cos^{-1}x\leq\pi$. The case of $t_y=t_1$ corresponds to the original Haldane model and the Weyl points $W_1$ (for $M=3\sqrt{3}t_2\sin\phi$) and $W_2$ (for $M=-3\sqrt{3}t_2\sin\phi$) are located at $K'$ and $K$ points of the Brillouin zone, respectively.

For $t_y=0$, the system is a Weyl semimetal in the range $-4t_2\abs{\sin\phi}\leq M \leq 4t_2\abs{\sin\phi}$, and there are two Weyl points in the reciprocal space at
\be
    W_3:\qty(k_x,k_y)&=\qty(\frac{\pi}{\sqrt{3}a},\frac{2}{3a}\cos^{-1}\qty(-\frac{M}{4t_2\sin\phi}))\label{eq:W3},\\
    W_4:\qty(k_x,k_y)&=\qty(\frac{\pi}{\sqrt{3}a},-\frac{2}{3a}\cos^{-1}\qty(-\frac{M}{4t_2\sin\phi}))\label{eq:W4},
\ee
where $\sin\phi\neq0$ is assumed. In particular, at $M=\pm4t_2\sin\phi$, $W_3$ and $W_4$ overlap, and they are equal to $W_1$ (for $M=4t_2\sin\phi$) or $W_2$ (for $M=-4t_2\sin\phi$).

The monopole charges of Weyl points $W_{1,2}$ in the $(\vb{k},M)$ space are calculated as 
\be
    Q_{1}&=-\mathrm{sgn}(t_y),\label{eq:Q1}\\
    Q_{2}&=\mathrm{sgn}(t_y),\label{eq:Q2}
\ee
respectively for $t_y\neq0$. For $t_y=0$, monopole charges in the $(\vb{k},t_y)$ space are
\be
    Q_3 = Q_4 = \mathrm{sgn}(\sin\phi)\label{eq:Q34}
\ee
for $W_3$ and $W_4$, respectively for $\sin\phi\neq0$. These values of the monopole charges are calculated analytically by expanding the $2\times2$ Hamiltonian to the linear order around the Weyl point by following Ref.~\cite{Murakami2008}. These Weyl points are shown in Figs.~\ref{fig:ExtHaldane}(b1) to (b5) for various values of $t_y/t_1$ and $M$ with $\sin\phi>0$ as an example.

\begin{figure}[t]
    \begin{center}
        \includegraphics[width=0.8\columnwidth,pagebox=artbox]{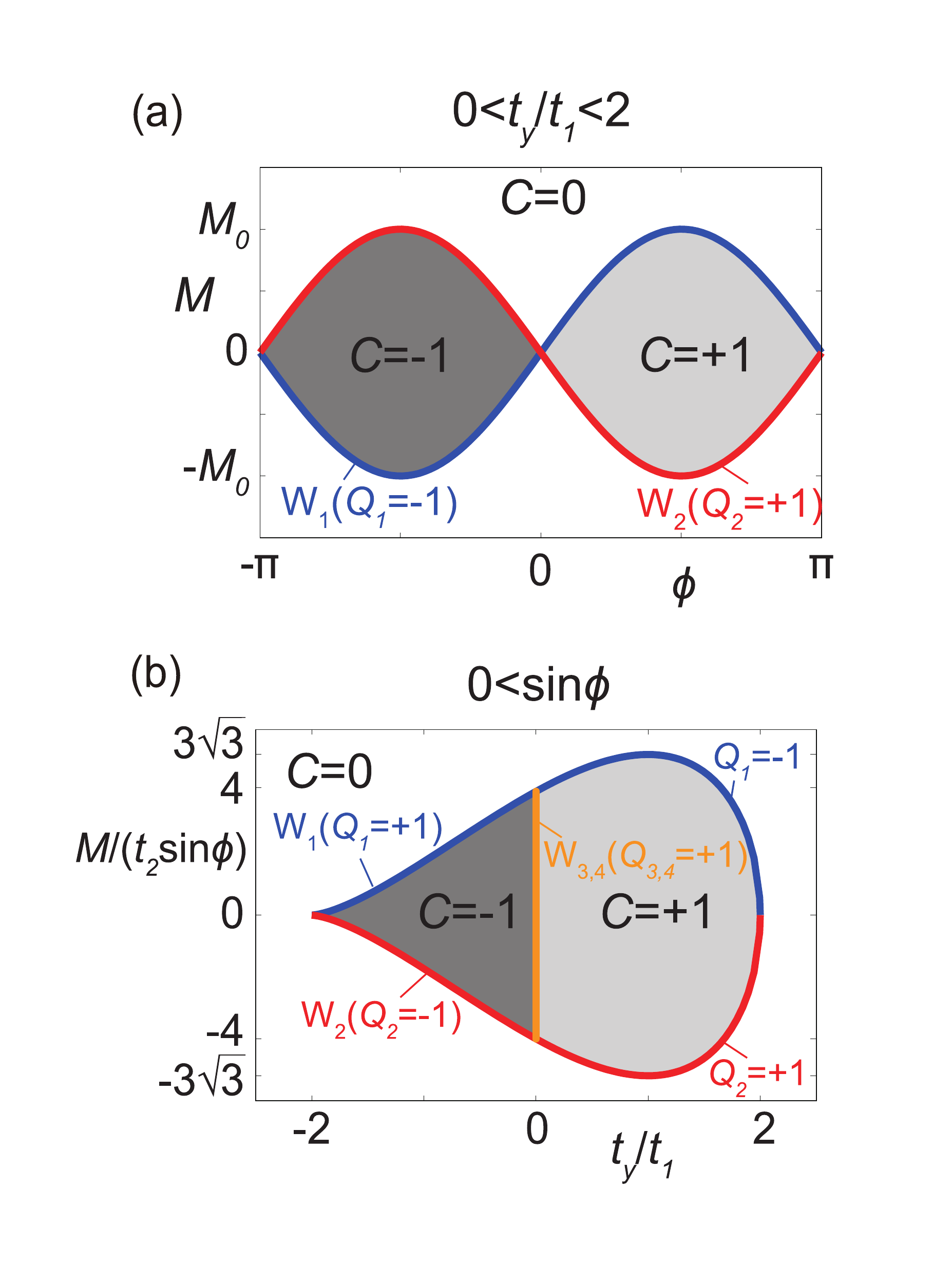}
        \caption{The phase diagrams of the extended Haldane model. (a) Phase diagram of the extended Haldane model in the $\qty(\phi,M)$ space for $t_y/t_1\in(0,2)$. The phase boundaries are $M=\pm M_0\sin\phi$. For $t_y/t_1\in(-2,0)$, the diagram becomes the inverted image with respect to the $\phi$ axis with signs of the Chern number also inverted. (b) Phase diagram in the $\qty(t_y/t_1,M/(t_2\sin\phi))$ space for $\sin\phi>0$. The phase boundaries are $M/(t_2\sin\phi)=\pm4\qty(1+t_y/(2t_1))\sqrt{1-(t_y/(2t_1))^2}$. For $\sin\phi<0$, the diagram becomes inverted with respect to the $t_y/t_1$ axis with signs of the Chern number also inverted.}
        \label{fig:ExtHaldane_Phase}
    \end{center}
\end{figure}

Similarly to the original Haldane model, the phase diagram of the extended Haldane model is shown in Fig.~\ref{fig:ExtHaldane_Phase}(a) for the parameter space $\qty(\phi,M)$. The regions enclosed by two sinusoidal curves are topologically nontrivial Chern insulator phases having non-zero Chern numbers and is not adiabatically connected to the topologically trivial region outside. A phase diagram can be drawn also for the parameter space $\qty(t_y/t_1,M/(t_2\sin\phi))$ as shown in Fig.~\ref{fig:ExtHaldane_Phase}(b) for $\sin\phi>0$. At the topological phase transitions with the change of the Chern number, a two-dimensional Weyl semimetal phase appears, and the change of the Chern number is equal to the sum of the monopole charges of the three-dimensional Weyl points in the $(\vb{k},M)$ or $(\vb{k},t_y)$ spaces at the transition. Using this model, we calculate the jump of electric polarization at the transition from the Chern insulator to the normal insulator below.

\subsection{Electric polarization in Chern insulators}
\label{subsec:RevPinChern}

Before discussing the jump of polarization, we need to know whether it is possible to define the electric polarization in the Chern insulator phase and if possible, how the definition differs from that of the normal insulator phase. It is a nontrivial issue since the edges are metallic. We here briefly review the paper by Coh and Vanderbilt~\cite{Coh2009} on the electric polarization in Chern insulators.

In the Chern insulator phase, the Berry phase is not a periodic function of the wavevector. The Berry phase defined by using a cell periodic part of the Bloch state $\ket{u}$ as
\be
    \phi_{2}(k_{1}) \coloneqq i\int_0^{b_{2}}\mathrm{d}k_{2} \bra{u}\pdv{}{k_{2}}\ket{u},
\ee
satisfies a relation
\be
    \phi_{2}(k_{0,1}+b_{1})-\phi_{2}(k_{0,1})=2\pi C,
\ee
where $C$ is a Chern number of the system and $k_{0,1}$ is a component along $\vb{b}_1$ of an arbitrary point $\vb{k}_0$ in the reciprocal space. Now, we define the electric polarization through an integration of this Berry phase as
\be
    \vb{P}^{[\vb{k}_0]}\coloneqq\frac{-ie}{(2\pi)^2}\int_{[\vb{k}_0]}\mathrm{d}^2 k\bra{u}\pdv{}{\vb{k}}\ket{u},\label{eq:CohVanderbilt_Chern}
\ee
where the integration is over a parallelogram with vertices $\vb{k}_0$, $\vb{k}_0+\vb{b}_1$, $\vb{k}_0+\vb{b}_1+\vb{b}_2$, and $\vb{k}_0+\vb{b}_2$ for a wavevector $\vb{k}_0$ and the Bloch wavefunction is taken to be continuous within this parallelogram. Then, $\vb{P}$ depends on the choice of $\vb{k}_0$ in the Chern insulator phase as
\be
    \vb{P}^{[\vb{k}_0+\Delta\vb{k}]}-\vb{P}^{[\vb{k}_0]}=-\frac{eC}{2\pi}\hat{z}\times\Delta\vb{k}.
\ee
The electric polarization is expected to be equal to the surface charge density $\sigma$ through the surface theorem $\sigma = \vb{P}^{[\vb{k}_0]}\cdot\vb{n}$, where $\vb{n}$ is a unit normal vector of the surface. Since $\sigma$ should not depend on $\vb{k}_0$, $\vb{P}^{[\vb{k}_0]}$ cannot be a physical polarization as it is. In other words, the value of $\vb{k}_0$ in $\vb{P}^{[\vb{k}_0]}$ has to be properly chosen to get a physical value of $\sigma$~\cite{Coh2009}.

\subsection{Definition of polarization jump between a Chern insulator and a normal insulator phase}
\label{subsec:Def_jump_Chern}

We now define the jump of polarization at phase transitions between a Chern and a normal insulator phase. The crucial difference from a transition between two normal insulator phases is that the Chern insulator phase cannot be adiabatically connected to the normal insulator phase. This makes the jump of polarization ill-defined in general. Namely, since $\vb{P}$ in the Chern insulator phase depends on $\vb{k}_0$ but is independent of $\vb{k}_0$ in the normal insulator phase, their difference also depends on the choice of $\vb{k}_0$. Instead of comparing $\vb{P}$ between the two phases, we introduce an extended polarization $\tilde{\vb{P}}$ within the Chern insulator phase and compare it with the polarization $\vb{P}$ in the normal insulator phase. Suppose that the system is in a Chern insulator phase for $M<M_0$ and is in a normal insulator phase for $M>M_0$. Then we define the extended polarization $\tilde{\vb{P}}$ as
\be
    \tilde{\vb{P}}(M) &\coloneqq \vb{P}^{[\vb{k}_0]}(M)\qquad(M<M_0),
\ee
where $\vb{k}_0$ is taken to be $\vb{P}^{[\vb{k}_0]}=0$ at an inversion symmetric or two-fold rotational symmetric point $\vb{k}_0=\frac{n}{2}\vb{b}_{1}+\frac{m}{2}\vb{b}_{2}\,(n,m\in\mathbb{Z})$ in the Chern insulator phase on the phase diagram. In inversion or two-fold rotational symmetric systems, one can then show $\vb{P}^{[\vb{k}_0]}\equiv\vb{0}$ by extending the proof for atomic insulators in the Supplement of Ref.~\cite{Benalcazar2019} to Chern insulators. In inversion or two-fold rotational symmetric systems, electric polarization is restricted to zero or half the polarization quantum in terms of modulo the polarization quantum $\frac{e}{\Omega}\vb{a}_{1,2}$. Thus, within this definition $\tilde{\vb{P}}(M)$ is defined in terms of modulo $\frac{e}{2\Omega}\vb{a}_{1,2}$.

Note that in general, we can take any inversion symmetric or two-fold rotational symmetric point as a reference provided that we can make $\vb{P}^{[\vb{k}_0]}=\vb{0}$ by choosing $\vb{k}_0$ appropriately. Then, the extended polarization $\tilde{\vb{P}}$ satisfies the surface theorem $\sigma = \tilde{\vb{P}}\cdot\vb{n}$ for the surface charge density $\sigma$ of an inversion symmetric or two-fold rotational symmetric finite system such as the ribbon-shaped system we introduce in Sec.~\ref{sec:origin}. On the other hand, we cannot use a three-fold rotational symmetric point, for example, as a reference since this symmetry is not preserved in a ribbon-shaped system. A discussion on the jump of polarization in systems without any such symmetries is given at the end of Sec.~\ref{sec:origin}. Similarly, there are choices of $\vb{k}_0$ which satisfy $\vb{P}^{[\vb{k}_0]}=\vb{0}$ but such choices have no impact on the resulting polarization. Then, $\tilde{\vb{P}}(M)$ is independent of our choice of reference and it is connected to the physical quantity $\sigma$. We can now define the jump of polarization at $M=M_0$ as
\be
    \Delta\vb{P} &\coloneqq \lim_{\delta\to0^+}\qty[\vb{P}(M_0+\delta)-\tilde{\vb{P}}(M_0-\delta)].\label{eq:DeltaP_Chern_gen}
\ee
In the extended Haldane model, $M=0$ is a line where the system becomes inversion symmetric. In inversion symmetric systems, electric polarization is restricted to zero or half the polarization quantum in terms of modulo the polarization quantum $\frac{e}{\Omega}\vb{a}_{1,2}$.

\begin{figure}[t]
    \begin{center}
        \includegraphics[width=\columnwidth,pagebox=artbox]{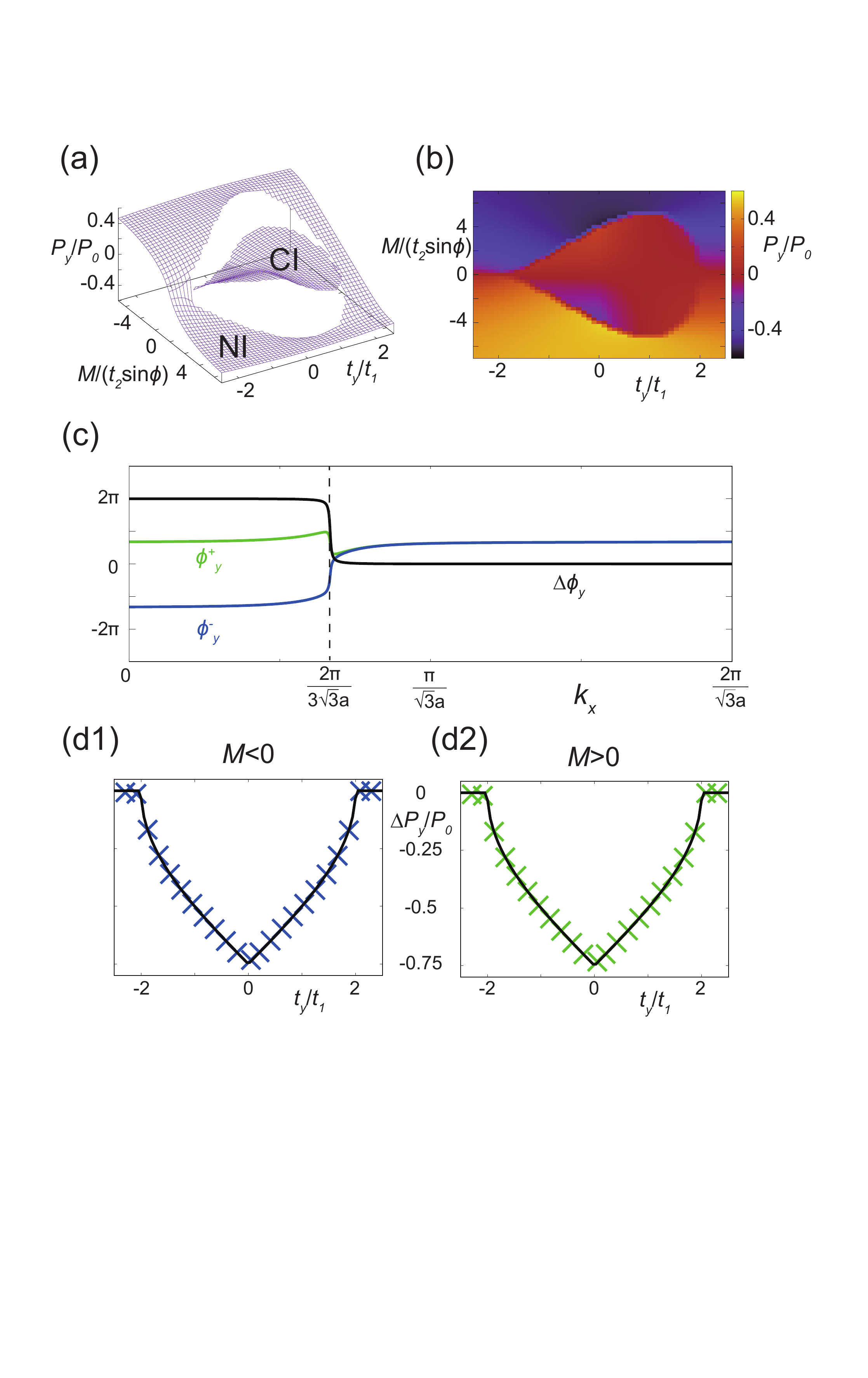}
        \caption{The $y$-component of the polarization of the extended Haldane model. (a) A three-dimensional plot of $P_{y}$ and $\tilde{P}_{y}$ scaled by $P_0\coloneqq\frac{ae}{\Omega}$ for $\sin\phi>0$ and (b) a two-dimensional color plot of the same $P_{y}$ in the normal insulator (NI) phase and $\tilde{P}_{y}$ in the Chern insulator (CI) phase. (c) Berry phases as a function of $k_x$ at the topological phase transition at $\phi=\frac{\pi}{2}$, $t_y/t_1=1$, and $M/t_2=3\sqrt{3}$. Two lines $\phi^{\pm}_y$ indicate Berry phases at $M/t_2=3\sqrt{3}\pm0.01$ and $\Delta\phi_y\coloneqq\phi^+_y-\phi^-_y$. (d1,2) The results of numerical calculations of jumps of electric polarization $\Delta P_y$ for (d1) $M>0$ and (d2) $M<0$ by markers. The analytical results from Eq.~\eqref{eq:DeltaP_Chern} are also plotted by black lines.}
        \label{fig:ExtHaldaneP}
    \end{center}
\end{figure}

Having defined $\Delta\vb{P}$ by Eq.~\eqref{eq:DeltaP_Chern_gen}, we confirm that $\Delta\vb{P}\neq\vb{0}$ at the phase transition using the extended Haldane model. We take $\vb{k}_0=\vb{0}$ and calculate the $y$-component of $\vb{P}$ and $\tilde{\vb{P}}$ in each phase as functions of $t_y/t_1$ and $M/(t_2\sin\phi)$. Results are plotted in Figs.~\ref{fig:ExtHaldaneP}(a) and (b). In this case, polarizations on the line $M=0$ for both normal and Chern insulator phases are zero. We can see that there are finite jumps at the phase boundaries shown in Fig.~\ref{fig:ExtHaldane_Phase}(b). Although there is a phase boundary at $t_y=0$ and $-4\leq M/(t_2\sin\phi)\leq4$ between Chern insulator phases with different Chern numbers, there is no jump of polarization there for $P_y$. We consider this point soon.

\subsection{Jump of polarization formula between a Chern insulator and a normal insulator phase}
\label{subsec_ChernFormula}

We next derive a general formula describing the jump of polarization at the phase transition between a Chern and a normal insulator phase. These jumps are caused by nonzero monopole charges of Weyl points as reviewed in Sec.~\ref{sec:RevPol}. As an example, we plot the Berry phases $\phi^{\pm}_y(k_x)$ in the extended Haldane model at $\phi=\frac{\pi}{2}, t_y/t_1=1$, and $M/t_2=3\sqrt{3}\pm0.01$, close to the phase transition $(M/t_2=3\sqrt{3})$ in Fig.~\ref{fig:ExtHaldaneP}(c). At $M/t_2=3\sqrt{3}$, a Weyl point $W_1$ with a monopole charge $Q_1=-1$ is present at the $K'$ point $\vb{k}^W=\qty(\frac{2\pi}{3\sqrt{3}a},\frac{2\pi}{3a})$. As expected from the general discussion in Sec.~\ref{sec:RevPol}, $\phi_y$ for $k_x$ larger and smaller than the projection of the Weyl point $k_x^W=\frac{2\pi}{3\sqrt{3}a}$ is expected to differ by $-2\pi$. Indeed, $\Delta\phi_y\coloneqq\phi^+_y-\phi^-_y$ plotted in Fig.~\ref{fig:ExtHaldaneP}(c) has a jump of $-2\pi$ at $k_x=\frac{2\pi}{3\sqrt{3}a}$. Since $\Delta\phi_y$ is constant at $k_x$ other than the projection of Weyl points and $\Delta P_y$ is given as an integration of $\Delta \phi_y$, the distance between $k_{0,x}$ and $k^W_x$ is needed to calculate $\Delta P_y$. Combining a similar result for $\phi_x$, we can write
\be
    \Delta \vb{P}=\frac{e}{2\pi}\sum_{W}Q^W\hat{z}\times\qty(\vb{k}^W-\vb{k}_0)\quad\qty(\mathrm{mod}\,\frac{e}{2\Omega}\vb{a}_{1,2}),\label{eq:DeltaP_Chern}
\ee
where $\vb{k}^W$ is the position of a Weyl point with a monopole charge $Q^W$, $\vb{k}_0$ is chosen using an inversion symmetric point as a reference, and the sum is taken over all Weyl points at the phase transition. From this derivation, the formula \eqref{eq:DeltaP_Chern} applies to any topological phase transitions with a change of the Chern number in general. Since two phases before and after the phase transition are not adiabatically connected, characterization of the jump of polarization by the accumulated current is vague. Nevertheless, this jump can still be observed experimentally through the surface charge density difference.

Now, we compare $\Delta \vb{P}$'s from our formula \eqref{eq:DeltaP_Chern} with the numerical results and see that they perfectly agree. At the phase transition from the Chern insulator to the normal insulator at $M = \pm M_0\sin\phi$, there is only one Weyl point $W_1$ ($M>0)$ or $W_2$ ($M<0$), whose position is given by Eqs.~\eqref{eq:W1} or \eqref{eq:W2} with monopole charges given in Eqs.~\eqref{eq:Q1} or \eqref{eq:Q2}. We plot the polarization jump $\Delta P_y$ from Eq.~\eqref{eq:DeltaP_Chern} by the black lines in Figs.~\ref{fig:ExtHaldaneP}(d1) and (d2) as functions of $t_y/t_1$, together with the numerical results of $\Delta P_y$ in Fig.~\ref{fig:ExtHaldaneP}(a) for the phase transitions at $M>0$ and $M<0$, respectively. The analytical results from Eq.~\eqref{eq:DeltaP_Chern} and numerical results $\Delta P_y$ show good agreements.

\begin{figure}[t]
    \begin{center}
        \includegraphics[width=.8\columnwidth,pagebox=artbox]{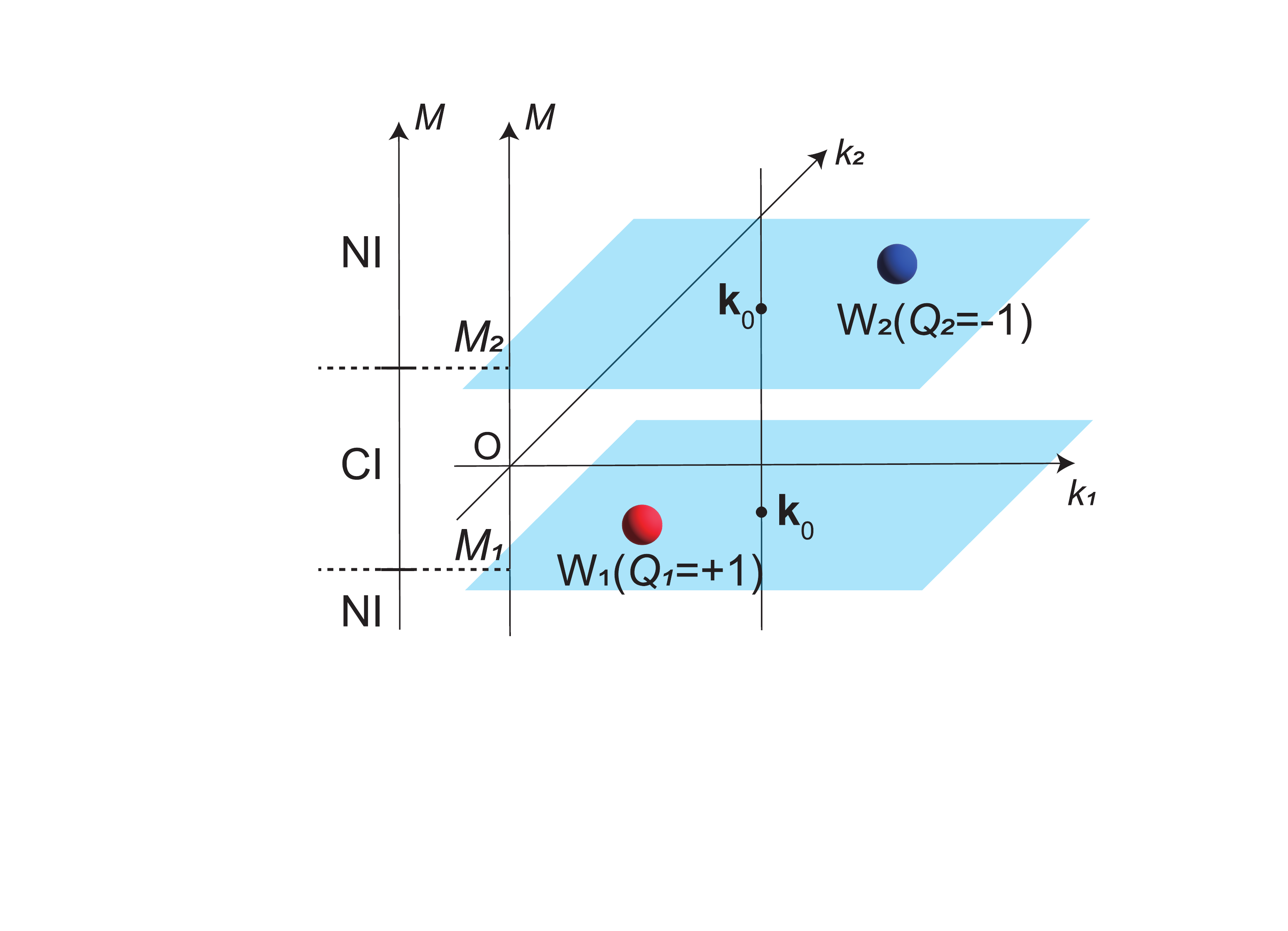}
        \caption{An example of Weyl-point distribution in the $(\vb{k},M)$ space in the reentrant transitions between the normal insulator (NI) and the Chern insulator (CI) phases. A Weyl point with monopole charge $Q_1=+1$ is located at $\qty(\vb{k}^{W_1},M_1)$ and another Weyl point with $Q_2=-1$ is located at $\qty(\vb{k}^{W_2},M_2)$. As $M$ is increased, the system changes from a normal insulator to a Chern insulator at $M=M_1$ and from a Chern insulator to a normal insulator at $M=M_2$.}
        \label{fig:3DkM_Chern}
    \end{center}
\end{figure}

We can graphically see that the result here is consistent with Ref.~\cite{Yoshida2023}. If the system has reentrant transitions between a normal insulator and a Chern insulator as we change a parameter $M$, Weyl points are separated in the direction of $M$ axis in the $(\vb{k},M)$ space as in Fig.~\ref{fig:3DkM_Chern}. When a Weyl point $W_1$ with monopole charge $Q_1=+1$ is located at $\qty(\vb{k}^{W_1},M_1)$ and another $W_2$ with $Q_2=-1$ is located at $\qty(\vb{k}^{W_2},M_2)$, the system is a Chern insulator in a region between two planes $M=M_1$ and $M=M_2$. In the limit where $M_1$ and $M_2$ become equal, the two topological phase transitions merge to become a transition between normal insulators. Thus, Fig.~\ref{fig:3DkM_Chern} approaches to Fig.~\ref{fig:3DkM_normal} and Eq.~\eqref{eq:DeltaP_Chern} reduces to Eq.~\eqref{eq:DeltaP_NI}.

We finally address the phase transition between two different Chern insulators with $C=+1$ and $-1$ at $t_y/t_1=0$ and $-4\leq M/(t_2\sin\phi)\leq4$. On this line, there are two Weyl points $W_3$ and $W_4$ at the positions given by Eqs.~\eqref{eq:W3} and \eqref{eq:W4} with the same monopole charges Eq.~\eqref{eq:Q34}. Then, from Eq.~\eqref{eq:DeltaP_Chern}, $\Delta \vb{P}$ is given as
\be
    \Delta\vb{P}&=\mqty(0\\\frac{3}{2}\mathrm{sgn}(\sin\phi)\\0)P_0\equiv0\quad\qty(\mathrm{mod}\frac{e}{2\Omega}\vb{a}_{1,2}),\label{eq:deltaP_example}
\ee
where $P_0=\frac{ae}{\Omega}$, $\vb{a}_1=a(\sqrt{3},0)$, and $\vb{a}_2=\frac{a}{2}(-\sqrt{3},3)$. It agrees with the absence of the jump across this phase transition in Fig.~\ref{fig:ExtHaldaneP}(a) and (b). Since there is a freedom in choosing different values of $\vb{k}_0$ at $M=0$ for $C=\pm1$, the jump of polarization is determined in terms of modulo a half the polarization quantum $\frac{e}{2\Omega}\vb{a}_{1,2}$ and not $\frac{e}{\Omega}\vb{a}_{1,2}$.

\section{\texorpdfstring{Polarization jump between $\mathbb{Z}_2$ topological insulator and normal insulator phase}{}}
\label{sec:Kane-Mele}

In this section, we construct a general theory of the polarization jump at phase transitions from a $\mathbb{Z}_2$-odd to a $\mathbb{Z}_2$-even phase. We first study the Kane-Mele model~\cite{Kane2005a} as an example. Similar to the Haldane model, the original Kane-Mele model possesses $C_3$ rotational symmetry, which prohibits the presence of electric polarization. We construct an extended Kane-Mele model by breaking this symmetry and then, we consider the jump of electric polarization between a $\mathbb{Z}_2$-odd and a $\mathbb{Z}_2$-even phase. We argue that this applies to general cases of $\mathbb{Z}_2$ topological phase transitions. Although this is a topological phase transition unlike the one in Ref.~\cite{Yoshida2023} reviewed in Sec.~\ref{sec:RevPol}, we find that the same description applies to the jump of polarization using Weyl dipoles.

\subsection{Extended Kane-Mele Model}
\label{subsec:ExtendedKM}

\begin{figure}[t]
    \begin{center}
        \includegraphics[width=\columnwidth,pagebox=artbox]{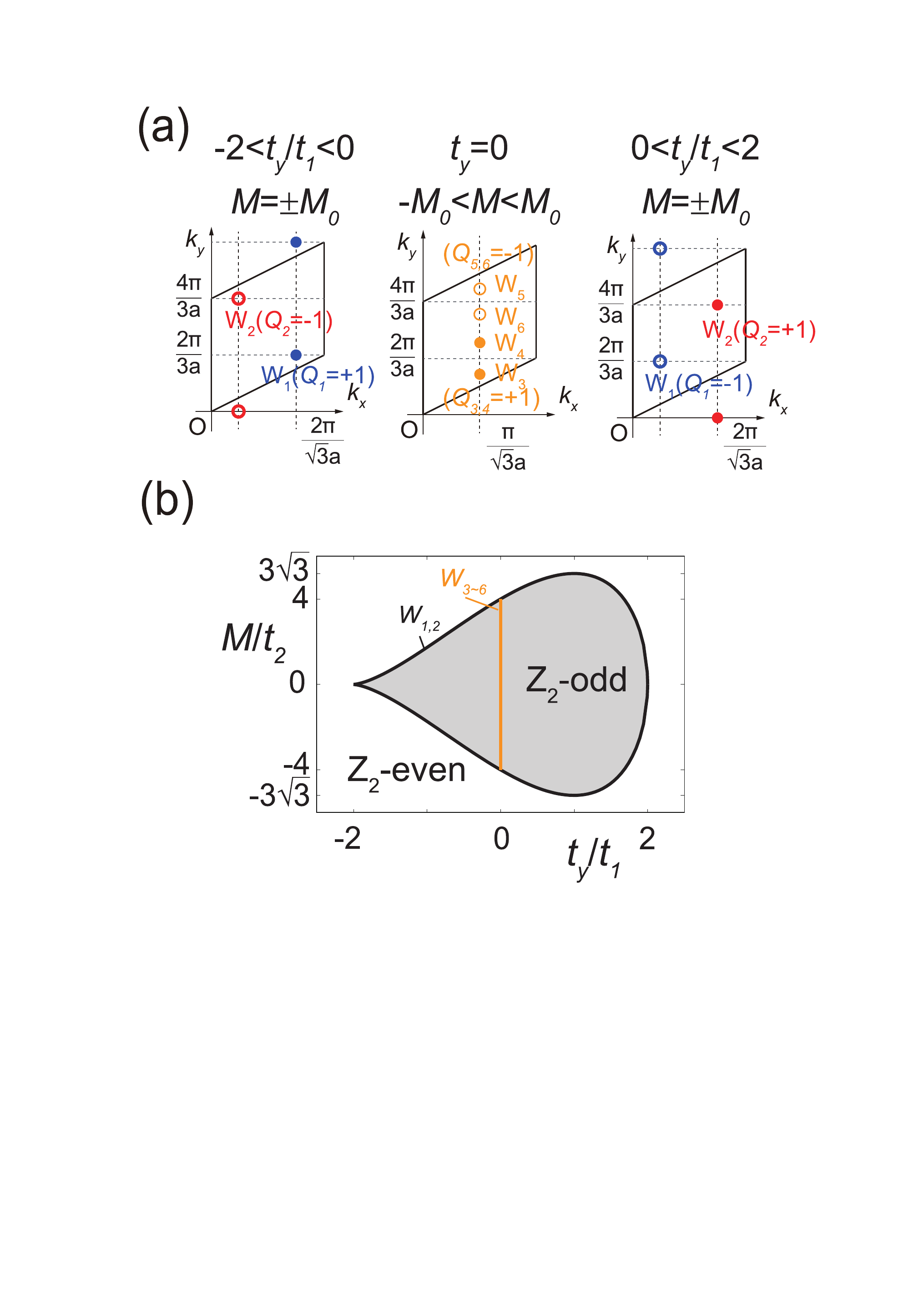}
        \caption{The extended Kane-Mele model without $C_3$ rotational symmetry. (a) Positions and monopole charges of Weyl points for indicated values of $t_y/t_1$ and $M$. Weyl points with opposite monopole charges appear in pairs and the sum of monopole charges inside the unit cell of the reciprocal lattice shown by the black parallelogram is always zero. (b) The phase diagram of the extended Kane-Mele model. The phase boundaries are given by $M/t_2=\pm M_0/t_2 = \pm4\qty(1+t_y/(2t_1))\sqrt{1-(t_y/(2t_1))^2}$.}
        \label{fig:ExtKM}
    \end{center}
\end{figure}

Since the ordinary Kane-Mele model without Rashba couplings is equivalent to the stacking of two Haldane models with $\phi=\pm\frac{\pi}{2}$, we construct the extended Kane-Mele model by stacking two extended Haldane models introduced in the previous section to break the $C_3$ rotational symmetry. We use the same notations as in Fig.~\ref{fig:ExtHaldane}(a) for NN hoppings, NNN hoppings, and on-site potentials. For simplicity, we neglect the Rashba couplings in the original Kane-Mele model here. Then, using $f(\vb{k})$ same as Eq.~\eqref{eq:f} and $g_2(\vb{k})\coloneqq-2t_2\sum_i\sin(\vb{k}\cdot\vb{d}_i)$, the Hamiltonian of this extended Kane-Mele model is given as
\be
    H(\vb{k}) = \Re f(\vb{k})\Gamma^1+\Im f(\vb{k}) \Gamma^{12}+M\Gamma^2+g_2(\vb{k})\Gamma^{15}, 
\ee
where we used the same representations of the Dirac matrices and their commutations as in Ref.~\cite{Kane2005a}. With Pauli matrices $\sigma^i$ and $s^i$ representing the sublattice and spin indices, they can be expressed as $\Gamma^1=\sigma^x\otimes I,\Gamma^2=\sigma^z\otimes I,\Gamma^{12}=\sigma^y\otimes I,\Gamma^{15}=\sigma^z\otimes s^z$. Four energy eigenvalues of this Hamiltonian are $E_{\pm,\pm}=\pm\sqrt{\abs{f}^2+(M\pm g_2)^2}$. At the half filling, this system becomes a Weyl semimetal by simultaneous gap closings at $W_1$ and $W_2$ located at $W_1=\qty(\frac{2}{\sqrt{3}a}\cos^{-1}\qty(\frac{t_y}{2t_1}),\frac{2\pi}{3a})$ and $W_2=\qty(\frac{2}{\sqrt{3}a}\cos^{-1}\qty(-\frac{t_y}{2t_1}),\frac{4\pi}{3a})$, which are the same positions as Eqs.~\eqref{eq:W1} and \eqref{eq:W2}, respectively if $t_y\neq0$ and $M= \pm M_0=\pm4t_2\qty(1+t_y/(2t_1))\sqrt{1-(t_y/(2t_1))^2}$. For $t_y=0$ and $-4\leq M/t_2\leq4$, there are four Weyl points
\be
    W_{3,4}&:\qty(\frac{\pi}{\sqrt{3}a},\pm\frac{2}{3a}\cos^{-1}\qty(-\frac{M}{4t_2})),\\
    W_{5,6}&:\qty(\frac{\pi}{\sqrt{3}a},\pm\frac{2}{3a}\cos^{-1}\qty(\frac{M}{4t_2})).
\ee

As this model is a stacking of two extended Haldane models, the monopole charges of the Weyl points are $Q_{1,2}=\mp\mathrm{sgn}(t_y)$ in the $(\vb{k},M)$ space, $Q_{3,4}=+1$, and $Q_{5,6}=-1$ in the $(\vb{k},t_y)$ space. These Weyl points appear in pairs with opposite signs of monopole charges as shown in Fig.~\ref{fig:ExtKM}(a) by the time-reversal symmetry, which guarantees that the Chern number is always zero. On the other hand, in the gray area shown in Fig.~\ref{fig:ExtKM}(b), the $\mathbb{Z}_2$ topological invariant defined in Ref.~\cite{Kane2005b} is odd while otherwise even. Hence, the lines $M=\pm M_0$ are phase boundaries between the $\mathbb{Z}_2$-odd phase and the $\mathbb{Z}_2$-even phase.

\subsection{\texorpdfstring{Electric polarization jump at the transition between a $\mathbb{Z}_2$-odd and a $\mathbb{Z}_2$-even phase}{}}
\label{subsec:PjumpZ2}

We next consider the electric polarization in the $\mathbb{Z}_2$-odd phase in general. In order for an electric polarization derived by Eq.~\eqref{eq:Polarization_KV} to have a physical meaning even in the $\mathbb{Z}_2$-odd phase, we naturally expect that the surface theorem $\sigma = \vb{P}\cdot\vb{n}$ is satisfied. The surface theorem in a normal insulator is proved through the Wannier representation of wavefunctions of electrons~\cite{Vanderbilt1993}. Hence, whether or not it is possible to construct well-localized Wannier functions is a crucial problem in the definition of polarization through Eq.~\eqref{eq:Polarization_KV}. While it is impossible to construct such Wannier functions in the Chern insulator phase~\cite{Thouless_1984}, it is possible in the $\mathbb{Z}_2$-odd phase by choosing an appropriate gauge~\cite{Soluyanov2011}. Therefore, we can use Eq.~\eqref{eq:Polarization_KV} without modification to calculate the electric polarization in the $\mathbb{Z}_2$-odd phase. However, since $\mathbb{Z}_2$-odd and $\mathbb{Z}_2$-even phases are not adiabatically connected, direct computation of the jump of polarization is impossible in $\mathbb{Z}_2$ topological insulators, either. Similar to the Chern insulator in Sec.~\ref{sec:Haldane}, we take a reference point within the $\mathbb{Z}_2$ topological insulator phase in the phase diagram, where we know the value of polarization from symmetries, and calculate the change of electric polarizations between a $\mathbb{Z}_2$-odd and a $\mathbb{Z}_2$-even phase.

\begin{figure}[t]
    \begin{center}
        \includegraphics[width=\columnwidth,pagebox=artbox]{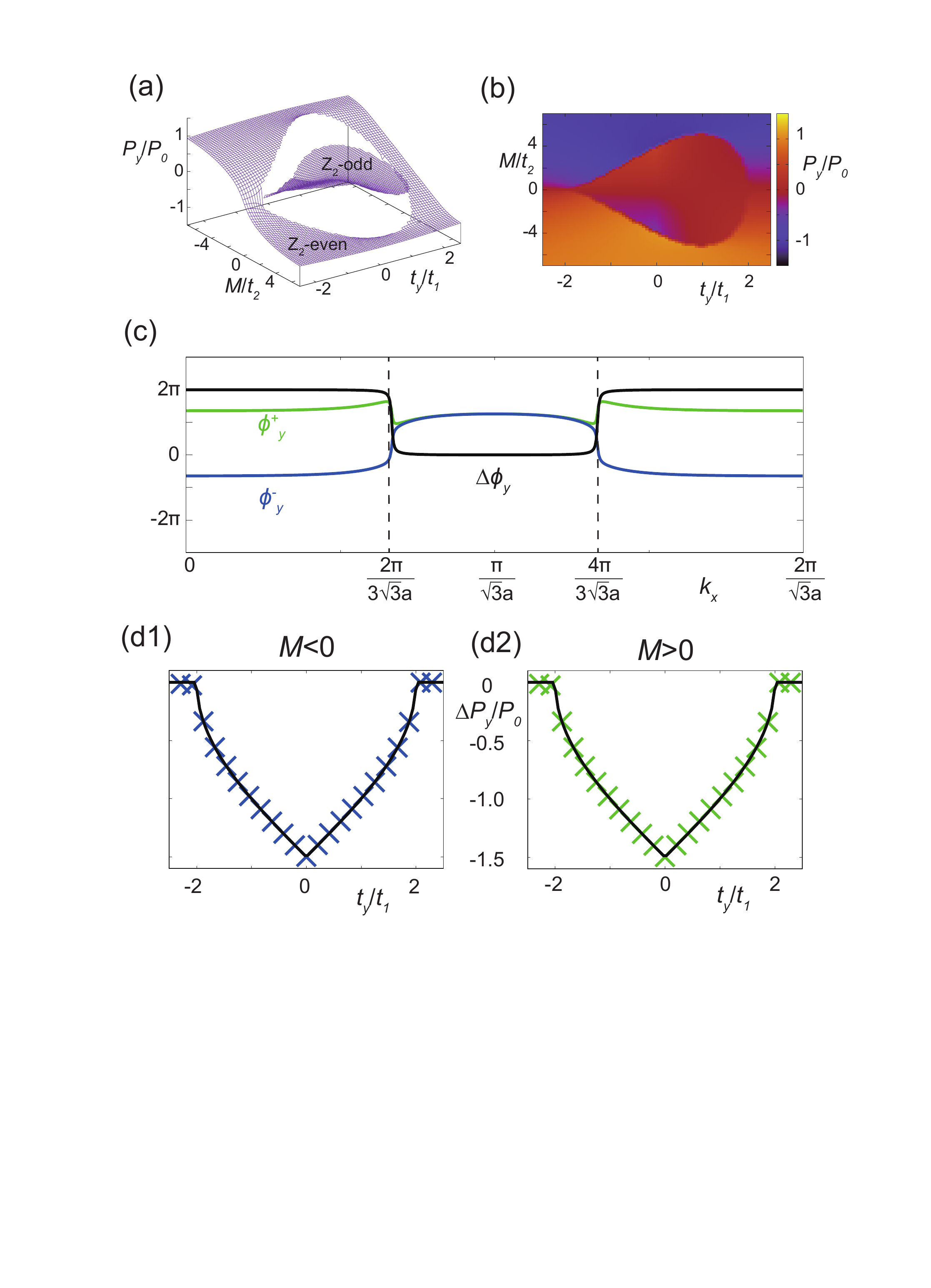}
        \caption{The polarization of the extended Kane-Mele model. (a) A 3D and (b) a 2D plot of $P_y$ in a parameter space $\qty(t_y/t_1,M/t_2)$ scaled by $P_0\coloneqq\frac{ae}{\Omega}$. (c) The Berry phases at the topological phase transition. The parameters are set to be $t_y/t_1=1$ and $M/t_2=3\sqrt{3}\pm0.01$ for $\phi^{\pm}_y$, respectively. The difference $\Delta \phi_y\coloneqq\phi^+_y-\phi^-_y$ is also plotted. (d1,2) The results of numerical calculation of jump of polarizations $\Delta P$ for (d1) $M<0$ and (d2) $M>0$, respectively are plotted by markers. The analytical results from Eq.~\eqref{eq:DeltaPy_Z2} are also shown by the black lines.}
        \label{fig:ExtKMP}
    \end{center}
\end{figure}

For the extended Kane-Mele model, $M=0$ is again a line where the system becomes inversion symmetric. Since the electric polarization in an inversion symmetric system is quantized to zero or half the polarization quantum in terms of modulo the polarization quantum $\frac{e}{\Omega}\vb{a}_{1,2}$, we can calculate the jump of polarization in terms of modulo $\frac{e}{2\Omega}\vb{a}_{1,2}$, which is half the polarization quantum. Figures~\ref{fig:ExtKMP}(a) and (b) show plots of the $y$-component of the polarization calculated by Eq.~\eqref{eq:Polarization_KV}. We can see that there are jumps of polarization at the phase boundary shown by the black curve in Fig.~\ref{fig:ExtKM}(b).

We next derive the formula of $\Delta\vb{P}$. As a function of $k_x$, the Berry phase difference $\Delta\phi_y$ across the phase transition jumps by $\pm2\pi$ at projections of Weyl points having monopole charges $\pm1$, respectively, as indicated in Fig.~\ref{fig:ExtKMP}(c). Different from the transition between a Chern insulator and a normal insulator in Sec.~\ref{sec:Haldane}, Weyl points with opposite monopole charges appear in pairs at $\vb{k}$ and $-\vb{k}$ in this case due to time-reversal symmetry. This is the same situation as the transition between normal insulators dealt with in Ref.~\cite{Yoshida2023}. Therefore, the jump of polarization can be described by using the Weyl dipole as
\be
    \Delta \vb{P} = \frac{e}{2\pi}\sum_W\qty(\hat{z}\times\vb{p}^W)\quad\qty(\mathrm{mod}\,\frac{e}{2\Omega}\vb{a}_{1,2}).\label{eq:DeltaP_Z2}
\ee
This argument is easily generalized to $\mathbb{Z}_2$ topological phase transitions in systems with time-reversal symmetry. In the extended Kane-Mele model, since the positions of the Weyl points $W_{1,2}$ at the $\mathbb{Z}_2$ topological phase transitions are given in Eqs.~\eqref{eq:W1} and \eqref{eq:W2}, we can calculate $\Delta P_y$ as a function of $t_y/t_1$ as
\be
    \Delta P_y \equiv \frac{3}{2\pi}\abs{\cos^{-1}\qty(\frac{t_y}{2t_1})-\cos^{-1}\qty(-\frac{t_y}{2t_1})}P_0.\label{eq:DeltaPy_Z2}
\ee
This analytical result is plotted by the black lines in Figs.~\ref{fig:ExtKMP}(d1) and (d2) together with numerical results, which show good agreements. We note that while $\vb{P}$ in the normal insulator phase is determined in terms of modulo the polarization quantum $\frac{e}{\Omega}\vb{a}_{1,2}$, $\vb{P}$ in the topological insulator phase has an ambiguity at reference points and it causes an ambiguity of $\Delta\vb{P}$ by a half of the polarization quantum. This is the significant point of the $\mathbb{Z}_2$ topological insulator compared with the normal insulator case. Although the jump is defined between two states that cannot be adiabatically connected, like in the case of the Chern insulator, this jump is experimentally measurable by surface charge density. 

Across the Weyl semimetal phase shown by the orange line in Fig.~\ref{fig:ExtKM}(b), where the gap closes in the $\mathbb{Z}_2$-odd phase, polarization changes continuously. Indeed, on this line, since the total Weyl dipole is given by $\sum_W\vb{p}^W=\qty[\frac{4\pi}{\sqrt{3}a},0]=2\vb{b}_1-\vb{b}_2\equiv0\ (\mathrm{mod}\,\vb{b}_{1,2})$, there is no jump across this line. We note that this absence or presence of a jump of polarization is not related to whether it is a topological phase transition or not, but rather, depending on the monopole charges and positions of Weyl points.

\section{Origin of the jump of polarization at topological phase transitions}
\label{sec:origin}

In this section, we clarify the origin of the jump of polarization at topological phase transitions from the view point of chiral or helical metallic states at the edge of the system in topological insulators. 

\begin{figure}[t]
    \begin{center}
        \includegraphics[width=\columnwidth,pagebox=artbox]{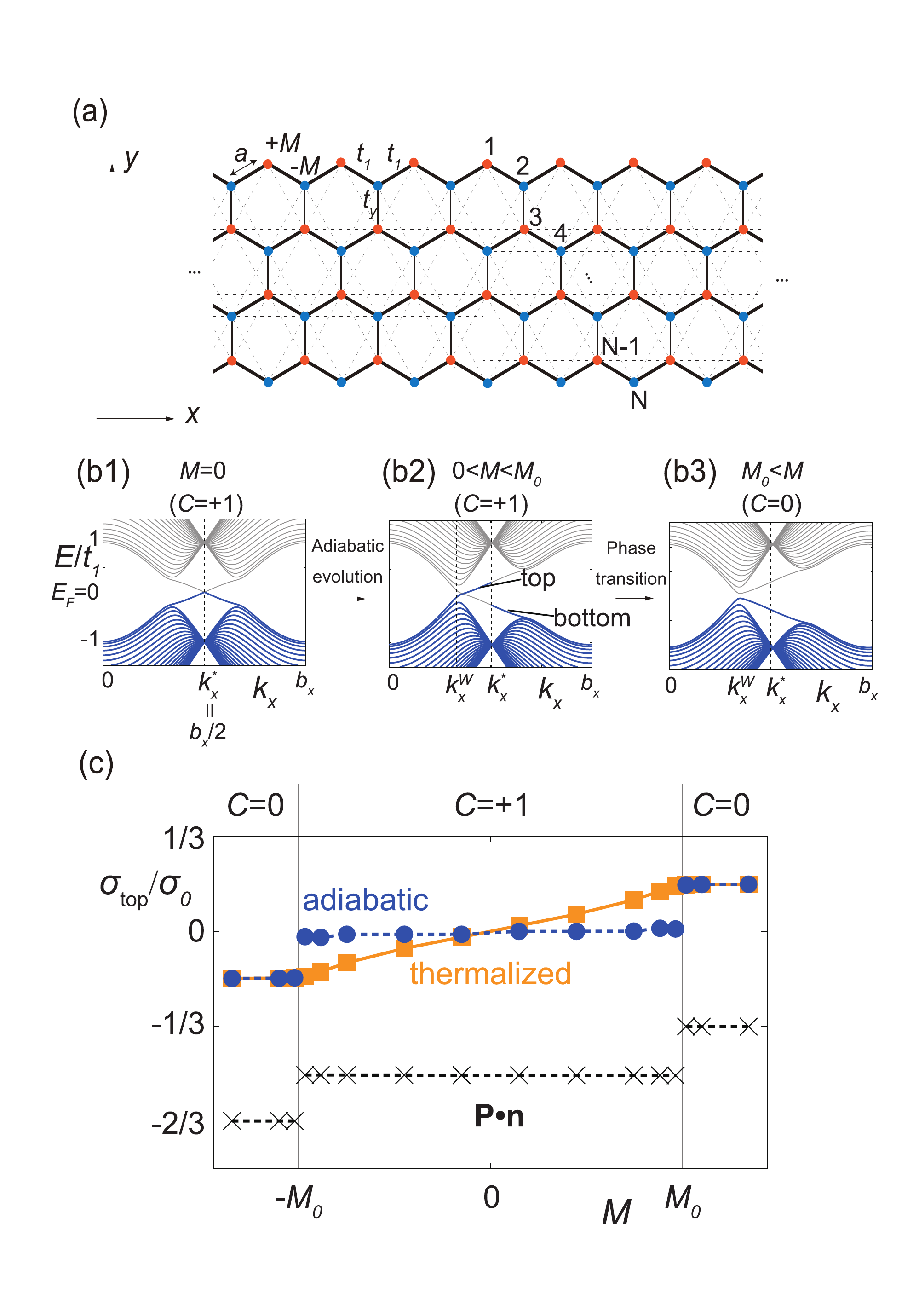}
        \caption{The ribbon geometry of the extended Haldane model. (a) The schematic illustration of the ribbon system extending infinitely in the $x$ direction and has $N$ sites in the $y$ direction. (b1)-(b3) Energy bands of the ribbon system of the extended Haldane model at (b1) $M=0$, (b2) $0<M<M_0$, and (b3) $M_0<M$. In the Chern insulator phase (b1) and (b2), we adopt the adiabatic filling, where blue bands are occupied by electrons. Occupation at the top and the bottom of the ribbon switches at $k_x^*=\frac{b_x}{2}$. (c) Edge charge densities at the top of the ribbon $\sigma_{\mathrm{top}}$ for thermalized and adiabatic fillings calculated by the window convolution method and the result from the modern theory of polarization for the bulk. In the calculation of $\vb{P}$, we take $\vb{k}_0=\qty(b_x/2,0)$. The unit of the charge density is $\sigma_0=e/(\sqrt{3}a)$ and the calculated value of $\vb{P}\cdot\vb{n}$ is defined modulo $\frac{\sigma_0}{2}$. Parameters are $t_y/t_1=1$ and $\phi=\frac{\pi}{2}$. The number of sites of the ribbon is $N=100$.}
        \label{fig:ExtH_ribbon}
    \end{center}
\end{figure}

We first review Ref.~\cite{Coh2009}, which states that to ensure $\sigma=\vb{P}\cdot\vb{n}$ in Chern insulators, we need to adopt the ``adiabatic'' filling as to how the chiral edge states of the Chern insulators are occupied. In this adiabatic filling, we first take a reference point in the phase diagram; in the present case, we take $M=0$ as a reference point where the inversion symmetry requires $\sigma=0$. This vanishing value of $\sigma$ is reproduced by assuming that all states below $E_F(=0)$ are occupied (Fig.~\ref{fig:ExtH_ribbon}(b1)). Let $k_x^*$ be a point where the occupation of edge states switches at $M=0$. In the case of the extended Haldane model, $k_x^*$ is given by $k_x^*=b_x/2$ as shown in Fig.~\ref{fig:ExtH_ribbon}(b1). In the adiabatic filling, as we change $M$, we assume that the value of $k_x^*$ stays the same as illustrated in Fig.~\ref{fig:ExtH_ribbon}(b2), where blue bands are occupied. This corresponds to the case where the evolution of $M$ is adiabatic, i.e. it is faster than the tunneling time between the top and bottom edges but slower than any other processes. Then, the edge charge density $\sigma$ defined as such satisfies $\sigma = \vb{P}^{[\vb{k}_0]}\cdot\vb{n}$, where the occupation switches in the ribbon geometry at $k^*_x$. We note that the adiabatic filling is different from the ``thermalized" filling, where the system is always thermalized as $M$ changes from zero and all electrons occupy bands below the Fermi energy $E_F$ for any value of $M$ in the Chern insulator phase. For the adiabatic filling in general, we can take $k_x^*$ or $k_y^*$ at the crossing point of edge states. If there are several crossing points, we can take one of them and the resulting $\sigma$ does not depend on our choice. If there are no crossing points of edge states, we cannot apply our theory.

To see the relationship between jumps of polarization and these edge states, we introduce ribbon-shaped systems which are infinitely long in the $x$ direction and have $N$ sites in the $y$ direction as in Fig.~\ref{fig:ExtH_ribbon}(a) for both extended Haldane and Kane-Mele models. We calculate the edge charge density $\sigma_{\mathrm{top}}$ at the top of the ribbon of the extended Haldane model as $\sigma_{\mathrm{top}}=\int_{y_{\mathrm{center}}}^{\infty}\mathrm{d}y \bar{\rho}(y)$, where $y_{\mathrm{center}}$ is the middle point of the ribbon. Here, the function $\bar{\rho}$ is defined in terms of the charge density $\rho(x,y)$ by the window convolution method~\cite{Baldereschi1988} as
\be
    \bar{\rho}(y)\coloneqq\frac{1}{bc}\int_{x_0}^{x_0+b}\mathrm{d}x\int_{y-\frac{c}{2}}^{y+\frac{c}{2}}\mathrm{d}y'\rho(x,y'),
\ee
where $b$ and $c$ are the lattice constants perpendicular to and along $\vb{n}$, respectively, $\vb{n}$ is the unit normal vector of the edges, and $x_0$ is an arbitrary value. We plot the edge charge density $\sigma_{\mathrm{top}}$ for thermalized and adiabatic fillings, together with the one calculated by the modern theory of polarization as $\sigma_{\mathrm{top}}=\vb{P}\cdot\vb{n}$, where $\vb{n}=(0,1)^{\mathrm{T}}$ for parameters $\phi=\frac{\pi}{2}$ and $t_y/t_1=1$ as a function of $M$ in Fig.~\ref{fig:ExtH_ribbon}(c). We take $\vb{k}_{0}=\qty(b_x/2,0)$ for the calculation of $\vb{P}$. We note that since $\vb{P}$ is determined in terms of modulo $\frac{e}{2\Omega}\vb{a}_{1,2}$, the calculated value of $\vb{P}\cdot\vb{n}$ is in terms of modulo $\frac{e}{2\sqrt{3}a}=\frac{\sigma_0}{2}$ where $\sigma_0=\frac{e}{\sqrt{3}a}$ is a unit of the edge charge density. In agreement with Ref.~\cite{Coh2009}, we can see that the modern theory of polarization well agrees with the adiabatic filling having a jump at $M=\pm M_0$ in terms of modulo $\frac{\sigma_0}{2}$ but not with the thermalized filling, which does not have a jump at $M=\pm M_0$. At the phase transition between a Chern and a normal insulator, the occupation of electrons at the edge state between $k_x^*$ and the Weyl point drastically changes in the adiabatic filling as shown in Figs.~\ref{fig:ExtH_ribbon}(b2) and (b3). At transition, $\frac{k_{0,x}-k_x^W}{2\pi}\qty(=\frac{k_{x}^*-k_x^W}{2\pi})$ electrons per unit length in $x$ direction moves from the top edge to the bottom edge, giving rise to the polarization jump by $\frac{e}{2\pi}(k_{0,x}-k^W_x)$, in agreement with Eq.~\eqref{eq:DeltaP_Chern}.

\begin{figure}[t]
    \begin{center}
        \includegraphics[width=0.9\columnwidth,pagebox=artbox]{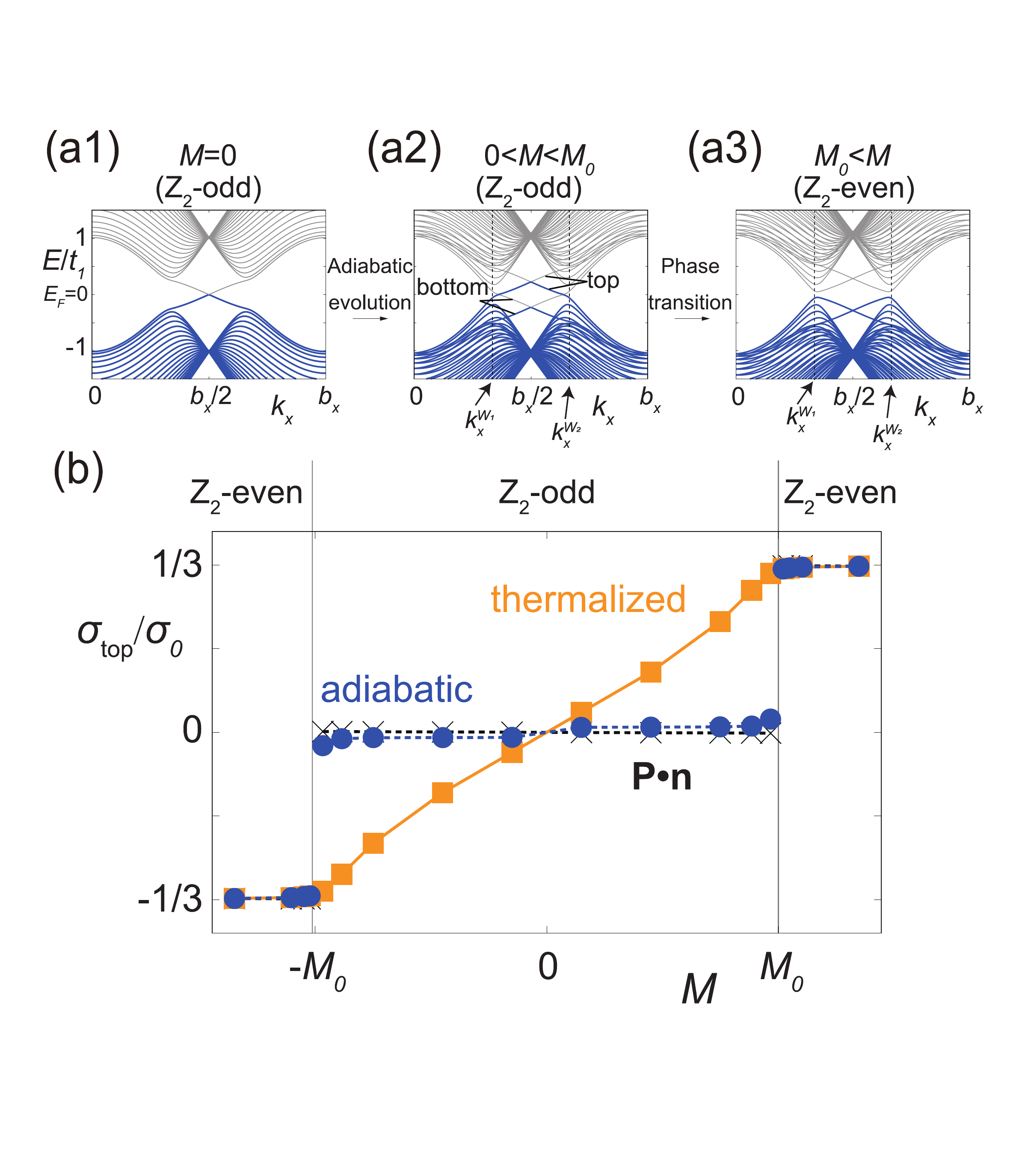}
        \caption{The ribbon geometry of the extended Kane-Mele model. (a1)-(a3) Energy bands of the ribbon system of the extended Kane-Mele model at (a1) $M=0$, (a2) $0<M<M_0$, and (a3) $M_0<M$. In the $\mathbb{Z}_2$-odd phase (a1) and (a2), we adopt the adiabatic filling, where blue bands are occupied by electrons. (b) Edge charge densities at the top of the ribbon $\sigma_{\mathrm{top}}$ for thermalized and adiabatic fillings calculated by the window convolution method and the result from the modern theory of polarization for the bulk. The unit of the charge density is $\sigma_0=e/(\sqrt{3}a)$. The parameter is set to be $t_y/t_1=1$ and the number of sites is $N=100$.}
        \label{fig:ExtKM_ribbon}
    \end{center}
\end{figure}

Similarly, we consider thermalized and adiabatic fillings for the ribbon geometry of the extended Kane-Mele model. In this case, the adiabatic evolution from $M=0$ (Fig.~\ref{fig:ExtKM_ribbon}(a1)) requires an inverted occupation of states between two crossing points of edge states as illustrated in Fig.~\ref{fig:ExtKM_ribbon}(a2) for the adiabatic filling, where blue bands are occupied. By the window convolution method, we calculate $\sigma_{\mathrm{top}}$'s for both fillings and compared it with $\vb{P}\cdot\vb{n}$ in Fig.~\ref{fig:ExtKM_ribbon}(b). We can see that in this case too, the modern theory of polarization gives a consistent result with the adiabatic filling. There is a drastic change of occupation at phase transitions as shown in Figs.~\ref{fig:ExtKM_ribbon}(a2) and (a3). This observation is consistent with Eq.~\eqref{eq:DeltaP_Z2}. Thus, the origin of jumps at topological phase transitions is the change of the occupation of electrons in the edge states. We note that in the case of the $\mathbb{Z}_2$ topological insulator phase, the bulk polarization is independent of the choice of high-symmetric reference point similar to the case of Chern insulators and it agrees with the surface charge density for the adiabatic filling.

To end this section, we consider a topological phase transition involving a change in the Chern number, specifically focusing on a case where the system has no such symmetries as to determine the reference point $\vb{k}_0$. If there are no points with inversion or two-fold rotational symmetries in the phase diagram, we cannot determine the $\vb{k}_0$ from the condition $\vb{P}^{[\vb{k}_0]}=0$ since the extended polarization does not satisfy the surface theorem for the surface charge density of those finite-size systems. In this case, we need to calculate $k_x^*$ and $k_y^*$ of the finite-size system in a geometry of our interest first. Then, $\vb{k}_0$ in the formula of the jump of polarization Eq.~\eqref{eq:DeltaP_Chern} is replaced by $\vb{k}^*\coloneqq\qty(k_x^*,k_y^*)$ as
\be
    \Delta \vb{P}=\frac{e}{2\pi}\sum_{W}Q^W\hat{z}\times\qty(\vb{k}^W-\vb{k}^*)\quad\qty(\mathrm{mod}\,\frac{e}{\Omega}\vb{a}_{1,2}),\label{eq:DeltaP_Chern_wo_sym}
\ee
by the discussion in Ref.~\cite{Coh2009} and the origin of the jump explained above. In contrast to the cases where we choose the reference point $\vb{k}_0$ to be inversion symmetric, as shown in Eqs.~\eqref{eq:DeltaP_Chern}, \eqref{eq:deltaP_example}, and \eqref{eq:DeltaP_Z2}, the jump of polarization is determined in terms of modulo the polarization quantum. Because of the existence of metallic edge states, we need information from the finite-size system in order to calculate the bulk polarization and its jump when symmetries cannot be used.



\section{Material proposal}
\label{sec:material}

\begin{figure}[t]
    \begin{center}
        \includegraphics[width=\columnwidth,pagebox=artbox]{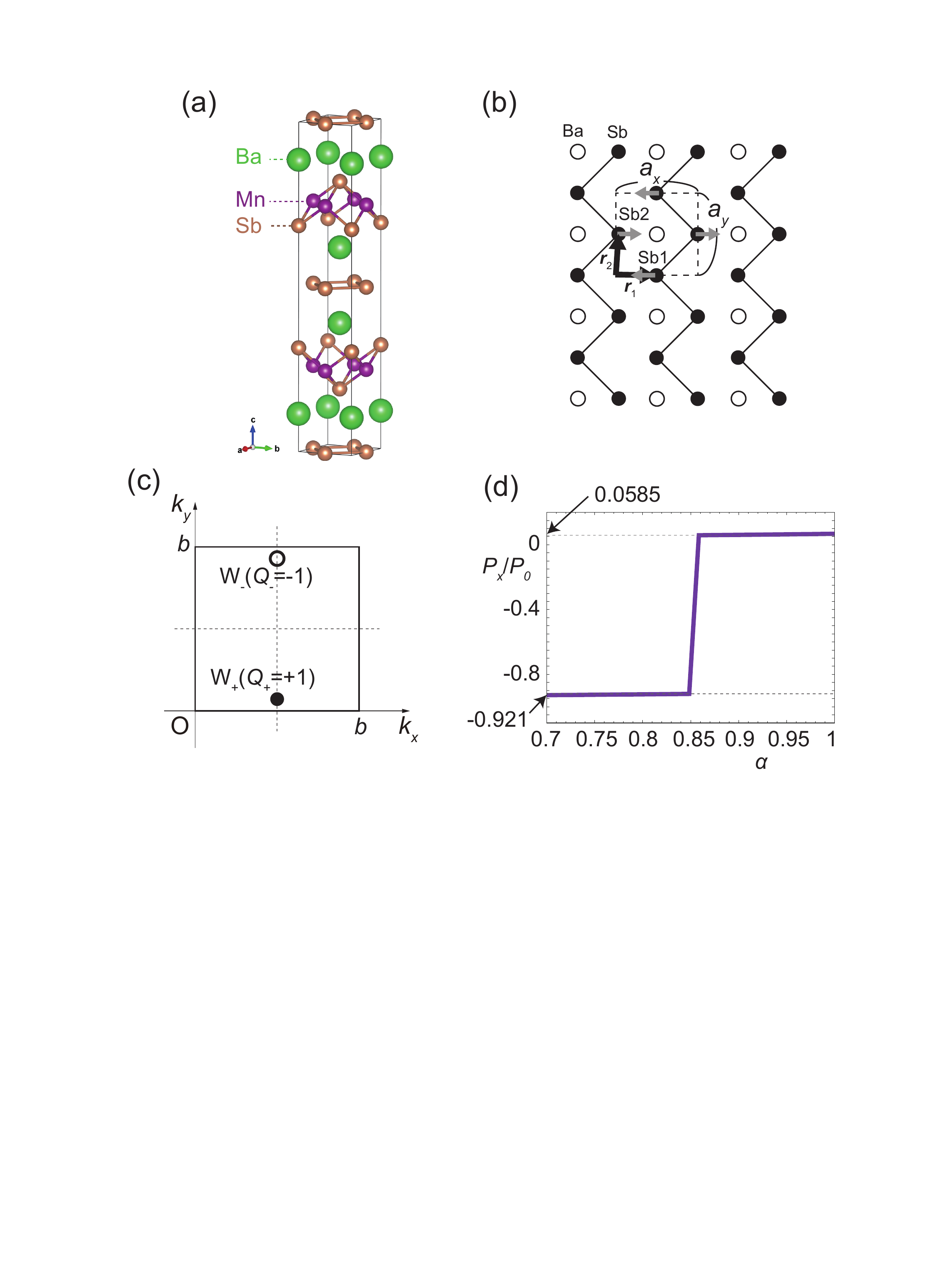}
        \caption{BaMnSb$_2$ as a candidate material. (a) A three-dimensional crystal structure of BaMnSb$_2$ by CrystalMaker. (b) A two-dimensional image of the Ba-Sb layer. The distortion shifts Sb atoms in the directions indicated by gray arrows. The dashed square is a unit cell with $a=a_x=a_y=4.5$ \AA. (c) Positions of Weyl points $W_{\pm}$ at $\alpha\approx0.86$, where $b=2\pi/a$ is the length of primitive reciprocal vectors. (d) The calculated $x$ component of the electric polarization. Here, $P_0=e/a$ is a unit of polarization of this system.} 
        \label{fig:BaMnSb2}
    \end{center}
\end{figure}

We here propose a material candidate to observe the jump of polarization at topological phase transitions. In this section, we apply our theory to a two-dimensional $\mathbb{Z}_2$ topological insulator $\mathrm{BaMnSb}_2$, which was studied in the context of a jump of the piezoelectric tensor in Ref.~\cite{Yu2020}. $\mathrm{BaMnSb}_2$ is a three-dimensional orthorhombic material ($I2mm$), which consists of alternately stacked Ba-Sb layers and Mn-Sb layers (Fig.~\ref{fig:BaMnSb2}(a)). By the insulating property of Mn-Sb layers, this system can be considered as a quasi-two-dimensional material whose conduction properties are determined by Ba-Sb layers as shown in Fig.~\ref{fig:BaMnSb2}(b)~\cite{Yu2020,sakai.etal2020}. By applying the distortion and shifting the positions of Sb atoms in the direction of arrows in Fig.~\ref{fig:BaMnSb2}(b), this system can be tuned to be either $\mathbb{Z}_2$-odd or $\mathbb{Z}_2$-even phases.

In the supplementary material of Ref.~\cite{Yu2020}, a tight-binding model based on the first-principle calculations is presented. The unit cell is a square with $a = a_x = a_y = 4.5$ \AA\ and unchanged by the distortion. The distortion is parametrized by $\alpha$, where $\alpha=0$ and $\alpha=1$ correspond to the non-distorted and fully-distorted cases, respectively. While the system is semimetallic up to $\alpha\approx0.4$, the system becomes a $\mathbb{Z}_2$-odd insulator by enhancing the distortion. Then, the phase transition from $\mathbb{Z}_2$-odd to $\mathbb{Z}_2$-even phase occurs at $\alpha\approx0.86$. At $\alpha\approx0.86$, two Weyl points $W_{\pm}$ are present at $\vb{k}_{\pm}=\qty(0.5,0.5\mp0.49)b$ as shown in Fig.~\ref{fig:BaMnSb2}(c), where $b=2\pi/a$ is the length of the primitive reciprocal vectors. Details of the tight-binding Hamiltonian and its parameters are given in the Appendix. From the effective Hamiltonian given in the supplementary material of Ref.~\cite{Yu2020}, monopole charges of these Weyl points at $\vb{k}_{\pm}$ are $Q_{\pm}=\pm1$, respectively. As a result, from Eq.~\eqref{eq:DeltaP_Z2}, the jump of the $x$ component of polarization with size $\Delta P_x\approx 0.98\frac{e}{a}$ is expected. Using the tight-binding Hamiltonian, we numerically calculated the $x$ component of the polarization around $\alpha=0.86$ in Fig.~\ref{fig:BaMnSb2}(d). We can see the jump of an expected size appears at the phase transition.

\section{Conclusion}
\label{sec:conclusion}

In this paper, we constructed a general theory on the behavior of electric polarization at topological phase transitions in two-dimensional systems. We used the extended Haldane model and the extended Kane-Mele model to study the jump of polarization at the phase transition from a Chern insulator to a normal insulator and from a $\mathbb{Z}_2$-odd to a $\mathbb{Z}_2$-even phase, respectively by tuning the on-site potential and anisotropic NN hopping parameters. Since topologically trivial and non-trivial phases are not adiabatically connected, we can define the jump of polarization between two phases by defining a reference point in the topological phase where we know the value of polarization by symmetry constraints.

In the Chern insulator phase, polarization depends on the corner of the integration $\vb{k}_0$ in the reciprocal space. By taking $\vb{k}_0$ to a value that satisfies the symmetry constraint at a reference point, we can define the polarization in the Chern insulator phase. At the phase transition where the Chern number changes, there appear Weyl points whose sum of monopole charges is not zero. In that case, the jump of polarization is expressed in terms of the positions of the Weyl points relative to $\vb{k}_0$. In the $\mathbb{Z}_2$ topological insulator phase, the definition of polarization is unchanged from that in the normal insulator phase since the Chern number is zero. In this case, Weyl points at a topological phase transition always appear in pairs with opposite monopole charges of $\pm Q$, and the jump of polarization is described by the Weyl dipole, which is originally introduced for the jump of polarization at transitions between normal insulator phases. We note that the presence or absence of particle-hole symmetry has no impact on our results, as exemplified by the fact that the extended Haldane model lacks particle-hole symmetry, while the extended Kane-Mele model possesses particle-hole symmetry.

At topological phase transitions, the change of the electron occupation in the edge states causes the jump of electric polarization. We also proposed BaMnSb$_2$ as a candidate material to observe the jump of polarization at topological phase transitions.

\begin{acknowledgments}
  This work is partly supported by Japan Society for the Promotion of Science (JSPS) KAKENHI Grant Numbers JP21K13865, JP22K18687, and JP22H00108.
\end{acknowledgments}

\appendix*
\section{\texorpdfstring{Tight-binding Hamiltonian of BaMnSb$_2$}{}}

In this appendix, we review the tight-binding Hamiltonian of BaMnSb$_2$ and its parameters used in Sec.~\ref{sec:material}, which are originally presented in the supplementary material of Ref.~\cite{Yu2020} for completeness. Details of the construction of the model are presented in the supplementary material of Ref.~\cite{Liu2021}.

As mentioned in Sec.~\ref{sec:material}, the Ba-Sb layer of BaMnSb$_2$ can be considered as a quasi-two-dimensional material. Main contributions to the bands near the Fermi energy are from $p_x$ and $p_y$ orbitals of Sb atoms and two Sb atoms are present in a unit cell as shown in Fig.~\ref{fig:BaMnSb2}(b), labeled as $1$ and $2$, that have sub-lattice vectors $\vb{r}_1=(x_1a,0)\coloneqq\qty((0.5-0.0488\alpha)a,0)$ and $\vb{r}_2=(x_2a,a/2)\coloneqq\qty(0.01729\alpha a,0.5a)$. Then, the bases of the tight-binding Hamiltonian are $\ket{\vb{R}+\vb{r}_i,\beta,s}$ with the lattice vector $\vb{R}=\qty(l_xa,l_ya)\,(l_{x,y}\in\mathbb{Z})$, the sublattice index $i=1,2$, the orbital index $\beta=p_x,p_y$, and the spin-$z$ index $s=\uparrow,\downarrow$. After the Fourier transformation, an $8\times8$ tight-binding Hamiltonian of this system $H(\vb{k})=h_0(\vb{k})+h_1(\vb{k})+h_2(\vb{k})$ consisting of the on-site term $h_0(\vb{k})$, the NN hopping term $h_1(\vb{k})$, and the NNN hopping term $h_2(\vb{k})$ is constructed. These terms are given by
\begin{widetext}
\be
    h_0(\vb{k})&=\mqty(M_1&\\&M_2),\\
    h_1(\vb{k})&=\mqty(0&0\\e^{-i[k_x(x_2-x_1)+k_y/2]a}&0)\otimes [T_1+T_2e^{-ik_xa}+T_3e^{-i(k_x-k_y)a}+T_4e^{ik_ya}]+h.c.,\\
    h_2(\vb{k})&=\mqty(Q_{x1}&\\&Q_{x2})e^{-ik_xa}+\mqty(Q_{y1}&\\&Q_{y2})e^{-ik_ya}+h.c..
\ee
The forms of $M$'s, $T$'s, and $Q$'s are
\be
    M_1 &=\tilde{m}_0\tau_0\sigma_0+\tilde{m}_1\tau_z\sigma_0+\lambda_0\tau_y\sigma_z,\ M_2 =C^{OS}_{4z}M_1\qty(C^{OS}_{4z})^{\dagger},\nonumber\\
    T_1 &=t_0\tau_0\sigma_0+t_1\tau_x\sigma_0+it_2\tau_y\sigma_0,\ T_2=\frac{\tau_z\sigma_yT_1\tau_z\sigma_y}{f(\alpha)},\nonumber\\
    T_4 &= \tau_z\sigma_yT_1\tau_z\sigma_y,\ T_3 = \frac{T_1}{f(\alpha)},\nonumber\\
    Q_{x1} &=t_3\tau_0\sigma_0+t_4\tau_z\sigma_0,\ Q_{x2}=t_5\tau_0\sigma_0+t_6\tau_z\sigma_0,\nonumber\\
    Q_{y1} &=C^{OS}_{4z}Q_{x2}\qty(C^{OS}_{4z})^{\dagger},\ Q_{y2}=C^{OS}_{4z}Q_{x1}\qty(C^{OS}_{4z})^{\dagger},
\ee
where $\tau$'s and $\sigma$'s are Pauli matrices for spin and orbital indices, $f(\alpha)=0.2\alpha+1$, and $C^{OS}_{4z}=-i\tau_ye^{-i\frac{\pi}{4}\sigma_z}$. Parameter values for numerical calculations are
\be
    \tilde{m}_0&=0,\ \tilde{m}_1=0.3\mathrm{eV},\ \lambda_0=0.25\mathrm{eV},\ t_0=1\mathrm{eV},\ t_1 = 2\mathrm{eV},\ t_2=0,\nonumber\\
    t_3&=0.1\mathrm{eV},\ t_4=-0.06\mathrm{eV},\ t_5=0.15\mathrm{eV},\ \mathrm{and}\ t_6=-0.06\mathrm{eV}.
\ee
\end{widetext}

\bibliography{PolTPT.bib}

\end{document}